\documentclass[aps,prb,twocolumn,showpacs,citeautoscript,superscriptaddress,longbibliography,eqsecnum]{revtex4-1}

\usepackage{verbatim}
\usepackage{amsmath}
\usepackage{amssymb}
\usepackage{colordvi}
\usepackage{graphicx}
\usepackage[unicode=true,pdfusetitle,
 bookmarks=true,bookmarksnumbered=false,bookmarksopen=false,
 breaklinks=false,pdfborder={0 0 1},backref=false,colorlinks=true,citecolor=blue]
 {hyperref}

\usepackage{soul}

\usepackage{mathtools}
\DeclarePairedDelimiter\floor{\lfloor}{\rfloor}



\begin{document}
\title{
Signatures of topological ground state degeneracy in Majorana islands}
 \author{Jukka~I.~V\"{a}yrynen}
\affiliation{Microsoft Quantum, Station Q, University of California, Santa Barbara, California 93106-6105, USA}

 \author{Adrian~E.~Feiguin}
\affiliation{Department of Physics, Northeastern University, Boston, Massachusetts 02115, USA}

 \author{Roman~M.~Lutchyn}
\affiliation{Microsoft Quantum, Station Q, University of California, Santa Barbara, California 93106-6105, USA}

\date{\today}

\begin{abstract}
We consider a mesoscopic superconducting island hosting multiple pairs of Majorana zero-energy modes. The Majorana island consists of multiple p-wave wires connected together by a trivial (s-wave) superconducting backbone and is characterized by an overall charging energy $E_C$; the wires are coupled to normal-metal leads via tunnel junctions. 
We calculate the average charge on the island as well as non-local conductance matrix as a function of a p-wave pairing gap $\Delta_P$, charging energy $E_C$ and dimensionless junction conductances $g_i$. We find that the presence of a topological ground-state degeneracy in the island dramatically enhances charge fluctuations and leads to the suppression of Coulomb blockade effects. 
In contrast with conventional (s-wave) mesoscopic superconducting islands, we find that Coulomb blockade oscillations of conductance are suppressed in Majorana islands regardless of the ratio $E_C/\Delta_P$ or the magnitude of the conductances $g_i$. We also discuss our findings in relation to the so-called topological Kondo effect. 
\end{abstract}

\maketitle

\section{Introduction}

Effect of quantum fluctuations in mesoscopic islands of superconducting metals has been extensively studied in the last two decades~\cite{grabert2013single,PhysRevLett.69.1997,PhysRevLett.70.1862,PhysRevLett.72.1742,eiles1994coulomb,2005PhRvB..72j4507H,von_Delft_2001}. In a conventional (s-wave) superconducting island connected to normal leads, the charge $\mathcal{Q}$ on the island varies in discrete steps as a function of an applied gate voltage $V_g$. When superconducting gap $\Delta$ is larger than the charging energy on the island $E_C$, the charge $\mathcal{Q}$ is $2e$-periodic and the ground-state is formed out of different number of Cooper pairs. Superconducting charge fluctuations smear out the transition between adjacent charge states, i.e.  $\mathcal{Q}$ and $\mathcal{Q}+2e$. In the limit of weak island-lead tunneling superconducting charge fluctuations are not strong enough to suppress Coulomb blockade effect~\cite{1993PhRvL..70.4138H,aleiner2002quantum}. Another way to probe properties of Coulomb-blockaded islands is to measure two-terminal tunneling conductance $G$. Away from charge degeneracy points (known as the ``valley'') tunneling of Cooper pairs across the superconducting island is suppressed at low temperatures $T$, {\it i.e.} $G \propto T^2$~\cite{PhysRevLett.119.116802,PhysRevB.99.014512}. 
This conclusion holds for any number of normal-metal (non-interacting) leads connected to the island.

The situation is different in the case of topological superconductors when an island hosts Majorana zero-energy modes (MZMs). The presence of these low-energy states drastically changes thermodynamic and transport properties~\cite{PhysRevLett.104.056402,2016PhRvB..93w5431V,2017PhRvL.119e7002L,2016PhRvB..94l5407L, 2012PhRvL.109o6803B,2013PhRvL.110u6803B,2013PhRvL.110s6401A,2014PhRvL.113g6401A,2014PhRvB..89d5143G,2014JPhA...47z5001A,2016PhRvB..94w5102H,2016arXiv160800581M,2017PhRvL.119b7701B,PhysRevB.97.235139}. This can be seen already in the case of two Majoranas  coupled to two ($M=2$)~\cite{PhysRevLett.104.056402,2016PhRvB..93w5431V,2017PhRvL.119e7002L,2016PhRvB..94l5407L} leads where resonant tunneling through a Majorana state (elastic cotunneling) dominates the conductance in the valley and leads to a finite contribution at $T=0$. 
Indeed, for $M=2$ one finds $G/ G_0  \sim (g\Delta_P / E_C)^2$ for $g\Delta_P  \ll E_C$, where $0\leq g\leq 1$ is the junctions' normal state dimensionless conductance, $\Delta_P$ is the topological pairing gap (i.e. p-wave gap), $E_C$ is the charging energy,  and $G_0 =  e^2/h$ is the conductance quantum. Thus, the strength of the superconducting charge fluctuations is controlled by the ratio $\Delta_P / E_C$ as well as the normal-state dimensionless conductance $g$.

The $M=2$ case is quite special because the charge on the island and fermion parity are locked by the charging energy removing the underlying ground-state degeneracy. In the $M$-terminal islands with $M>2$ MZMs charging energy fixes the overall fermion parity sector but it does not remove ground-state degeneracy, i.e. Majorana degrees of freedom form a SO$(M)$ impurity ``spin". It was proposed in Refs.~\onlinecite{PluggeSurface,Vijay2016,2017NJPh...19a2001P,2017PhRvB..95w5305K} that the remaining topological ground-state degeneracy can be used for quantum information processing. When such an island is coupled to normal leads, one can show that the topological ground-state degeneracy manifest itself in a number of different ways: the superconducting charge fluctuations are enhanced suppressing Coulomb blockade effect; the conductance $G_{ij}$ between leads $i$ and $j$ reaches a universal value $G_{i\neq j} = 2e^2/(Mh)$ at $T=0$ independent of the applied gate voltage.

In this paper we show that 
the suppression of Coulomb blockade effect can be used as a signature of topological ground state degeneracy. In order to draw general conclusions, we developed theoretical framework which is suitable for arbitrary ratio of $\Delta_P$ and $E_C$ and treats superconducting and charge fluctuations on equal footing. Our microscopic model for Majorana islands allows one to calculate observable quantities of interest (e.g. charge on the island and multiterminal conductance matrix) in terms of measurable parameters such as normal-state junction conductance $g$, $E_C$, $\Delta_P$ and applied dimensionless gate voltage $N_g$. We note that previous works~\cite{2012PhRvL.109o6803B,2013PhRvL.110u6803B,2013PhRvL.110s6401A,2014PhRvL.113g6401A,2014PhRvB..89d5143G,2014JPhA...47z5001A,2016PhRvB..94w5102H,2016arXiv160800581M,2017PhRvL.119b7701B,PhysRevB.97.235139} have considered the limit of $\Delta_P \rightarrow \infty$ in weak tunneling regime. Using existing experimental data on semiconductor-based Majorana islands~\cite{lutchyn2018majorana,2016Natur.531..206A, vaitieknas2018fluxinduced} one can set an upper bound on topological gap of $50\mu$eV indicating that $\Delta_P/E_C \ll 1$ regime is more experimentally relevant.

Our work can be placed in the context of the so-called topological Kondo effect~\cite{2012PhRvL.109o6803B,2013PhRvL.110u6803B,2013PhRvL.110s6401A,2014PhRvL.113g6401A,2014PhRvB..89d5143G,2014JPhA...47z5001A,2016PhRvB..94w5102H,2016arXiv160800581M,2017PhRvL.119b7701B,PhysRevB.97.235139}. 
In this exotic Kondo effect, the Coulomb  island of $M$ MZMs forms the  impurity ``spin'' with a $2^{\frac{1}{2}M-1}$-fold degenerate ground state in the Coulomb valley (the MZMs come in pairs which makes $M$ even). 
The impurity spin is screened by $M$ normal metal leads, each one tunnel-coupled to a different MZM. 
The topological Kondo effect drives the system to a strongly-coupled low-temperature fixed point where the conductance $G_{ij}$ 
reaches the  universal value quoted above. 
The finite-temperature corrections $\delta G_{ij}$ to this result vanish with  a universal non-Fermi-liquid power law exponent: $\delta G_{ij} \propto T^{2(1-\frac{2}{M})} $. 
The scaling of the non-local conductance with temperature can be used to detect Majorana-induced ground-state degeneracy in the island. Furthermore, multi-terminal conductance depends on the number of attached leads $M$ and cutting off one of the leads (by tuning the tunnel-gate) should change the conductance (i.e. $M \to M-1$). Note that once $M=2$ the system would cease to flow to the topological Kondo fixed point and conductance will become dependent on $\Delta_P /E_C$ as explained above. 

The structure of this paper is as follows: in Sec.~\ref{sec:qualit} we begin with the qualitative discussion of main  results. 
In Sec.~\ref{sec:model} we introduce   the microscopic model which is then mapped to a low-energy model of quantum Brownian motion (QBM). 
In Secs.~\ref{sec:Strong-superconductor-limit}--\ref{sec:Weak-superconductor-limit} we perform renormalization group analysis of the QBM model in the respective limits of large ($\Delta_P \gg E_C$) and small ($\Delta_P \ll E_C$) topological gap. 
By using the identified leading relevant or irrelevant operators and their scaling dimensions, we then proceed to evaluate the average charge and the conductance of the island in Sec.~\ref{subsec:StrongSCAverage-charge-and}. 
We compare our model to the ones studied in the context of the multi-channel Kondo problem in Sec.~\ref{sec:compare} and then draw conclusions in Sec.~\ref{sec:Conclusions}.

\section{Qualitative discussion of the main results \label{sec:qualit}}

\begin{figure}[tb]
\includegraphics[width=0.8\columnwidth]{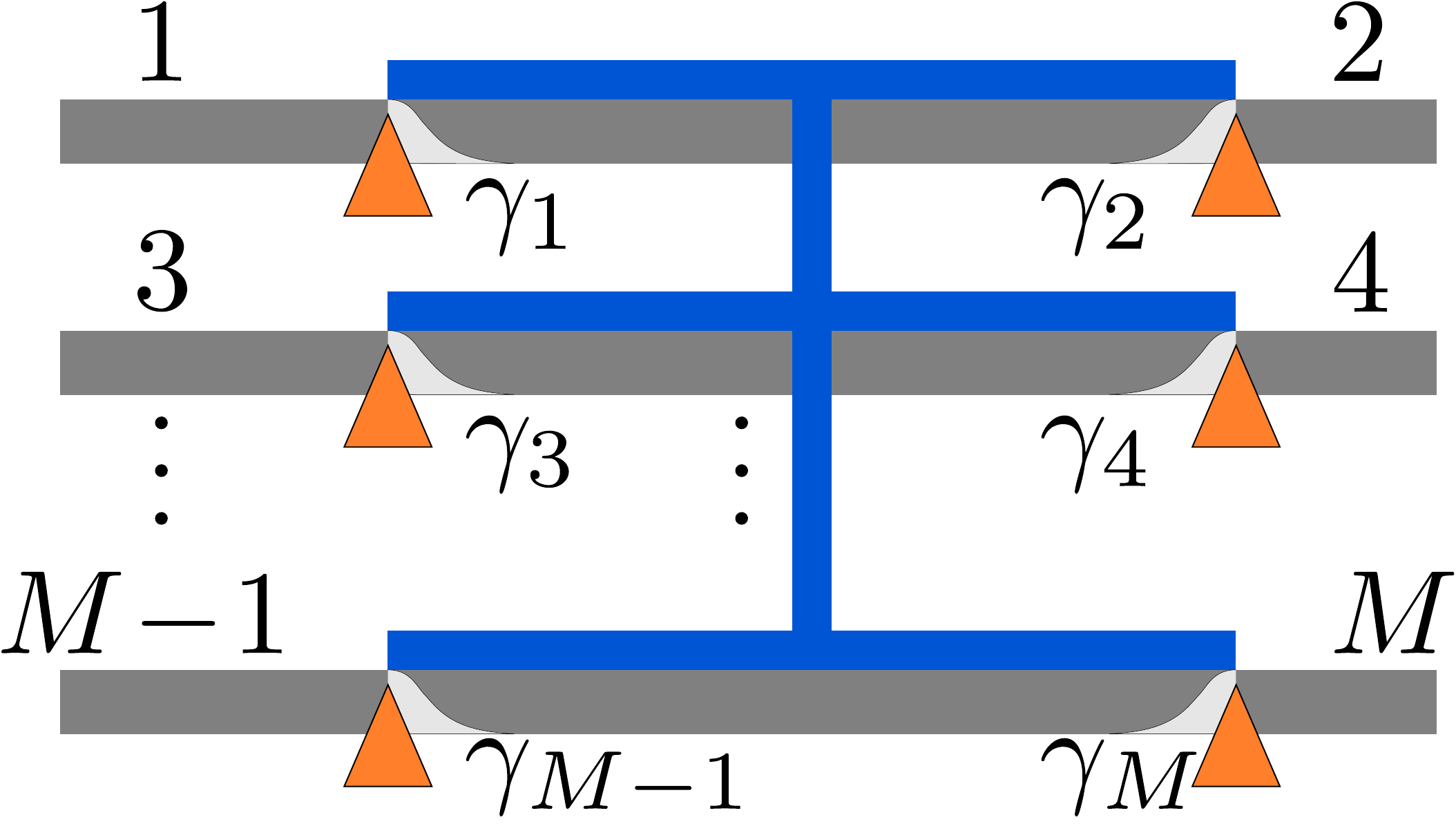}\caption{Schematic diagram of the device build out of $M/2$ nanowires. Each wire (grey) is split by two barriers (orange triangles) into left and right ``leads''  and a middle section. 
The middle section has a topological gap $\Delta_P$ due to a proximity  to  a central floating superconductor (dark blue). 
Well-separated Majorana zero modes (light grey) $\gamma_i$ are localized at the boundaries between the normal  and the topological superconducting regions. 
The middle section consisting of the nanowires and the s-wave superconductor has a  charging energy $E_C$. 
We assume that the  s-wave gap $\Delta_S$ of the central superconductor  is the largest energy scale in the problem, $\Delta_S \gg E_C, \Delta_P$. 
\label{fig:device}}
\end{figure}
\begin{figure}[tb]
\includegraphics[width=1.0\columnwidth]{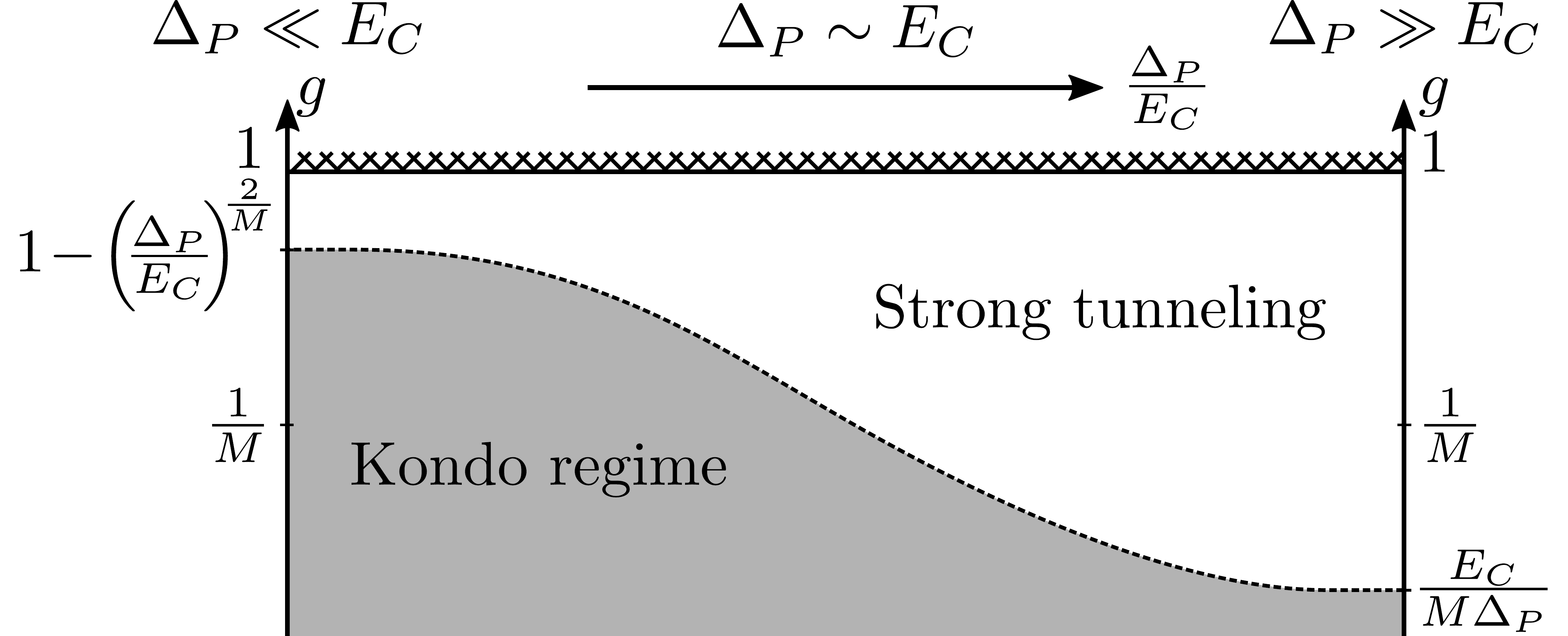}
\caption{
A diagram showing the behavior of the island at intermediate temperatures, before reaching the strong coupling fixed point. 
The parameter space is formed by the  dimensionless bare conductance  $0 \leq g \leq 1$ and the topological gap in units of charging energy $\Delta_P / E_C$. 
In the weak-coupling Kondo regime (grey)  there is a large temperature window $T_K \ll T \ll \min(E_{C},\,\Delta_{P})$ which is characterized by a marginally relevant tunneling operators. 
The strong tunneling regime (white) can be understood as the limit $T_K \to  \min(E_{C},\,\Delta_{P}) $. 
In this regime there is no logarithmic Kondo renormalization of tunneling amplitudes. Instead, the system is characterized by  weak irrelevant  reflection amplitudes that display  non-Fermi-liquid-like temperature-dependence at $T \ll \min(E_{C},\,\Delta_{P})$. 
Generally, the Kondo regime   exists when the renormalized tunneling amplitude  is small at scale $\min(E_{C},\,\Delta_{P})$. 
On the left, the charging energy is large, $ E_C \gg \Delta_P$, and even a  large bare tunneling amplitude (but still $1-g \gtrsim (\frac{\Delta_P}{E_C})^{(2/M)}$) gets renormalized to a small value at scale $\Delta_P$. 
On the right, the topological superconducting gap is large, $\Delta_P \gg E_C$, and even a small  tunneling amplitude (but still $g \gtrsim \frac{E_C }{M \Delta_P}$) gets renormalized to a large value at scale $E_C$. 
Thus, on the right, $\Delta_P \gg E_C$,  the Kondo regime only exists where bare conductance is  relatively small. The crossover between different regimes as a function of $\Delta_P/E_C$ and $g$ is schematically  indicated by the dashed line. 
}
\label{fig:gdiagram}
\end{figure}

We consider a Majorana island shown in Fig.~\ref{fig:device}, consisting of an  array of proximitized semiconductor nanowires tuned (by magnetic field and/or gate voltages) to the topological regime with the corresponding p-wave gap $\Delta_P$. In the regime of interest, the left and right halves of these wires form  $M$ effectively spinless semi-infinite leads. 
A central segment of length $L$ is separated from the leads by barriers which are characterized by a dimensionless conductance $g_i \in [0,1]$. 
We assume that $L$ is much larger than the p-wave coherence length so that we can ignore the hybridization of the Majorana states through the central segment. 
Proximitized nanowires are coupled with each other by trivial floating superconductor with the gap $\Delta_S \gg \Delta_P, E_C$. Different proximitized nanowires are separated from each other by the distance much larger than the coherence length $\xi_S$ in the trivial superconductor so that inter-wire Majorana hybridization can be neglected. 
The s-wave superconductor has a large number of channels connecting different segments of the device and therefore  the island formed by the superconductor together with the nanowires can be characterized by a single charging energy $E_C$ and inter-wire mutual charging effects can be neglected. 
For more details on the model, see Sec.~\ref{sec:model}. 

In this paper we  focus on four main parameter regimes summarized in Fig.~\ref{fig:gdiagram}. 
These regimes correspond to the cases of large or small $\Delta_P/ E_C$, and weak or strong tunneling, which is controlled by the dimensionless conductance $g$ of a junction. 
In all regimes the system flows to the universal strong coupling fixed point mentioned above. 
As a result, at low temperatures $T \ll T_K$, the conductance displays non-Fermi-liquid  corrections with a characteristic strong-coupling temperature scale $T_K$, see Eq.~(\ref{eq:SummaryGBelowTk}). 
This characteristic scale depends on the bare parameters and is summarized in Table~\ref{table}.  

In the ``Kondo regime'' in Fig.~\ref{fig:gdiagram}, the scale $T_K$ is well below both $E_C$ and $\Delta_P$. 
In this regime there is a wide temperature-window in which the junction conductances are small and exhibit the typical weak-coupling Kondo renormalization with logarithmic temperature-dependence.  
As seen in Fig.~\ref{fig:gdiagram}, this case requires small enough bare conductances $g$ and is  favored by small ratio $\Delta_P/E_C$.  
The Kondo temperature in this case can be written as, 
\begin{equation}
T_{K} \sim D_0 \exp\left[-\frac{1}{(M-2)\lambda(D_0)}\right]\,, \label{eq:SummaryTk}
\end{equation}
where $D_0 = \min(E_{C},\,\Delta_{P})$ sets the effective UV cutoff for the Kondo regime, and $\lambda(D_0)$ is the cotunneling amplitude, see Table~\ref{table}. 
This expression is valid in the Coulomb valley where $N_g$ is away from half-integer values. 

If the bare conductances are not too small, the scale $T_K$ becomes of order $\min(E_{C},\,\Delta_{P})$ [$\lambda(D_0) \sim 1/M$ in Eq.~(\ref{eq:SummaryTk}), see Table~\ref{table}]. 
This regime is denoted ``Strong tunneling'' in Fig.~\ref{fig:gdiagram} and is favored by a large ratio  $\Delta_P/E_C$. 
Benefiting from 
the large $T_K$, 
this  regime  thus may be advantageous for observing the non-Fermi-liquid aspects of the strong coupling fixed point.

In the rest of this section, we discuss thermodynamic and transport signatures of the topological ground-state degeneracy and summarize our findings for two observable quantities - the average charge on the island and the conductance matrix $G_{ij}$. 
For the latter, we focus on low temperatures below the strong coupling scale, $T \ll T_K$, where the results are universal, independent of the ratio $\Delta_P/ E_C$. 

\begin{table}
\begin{tabular}{|c|c||c|}
\hline 
\multicolumn{2}{|c||}{$E_{C}\gg\Delta_{P}$} & $E_{C}\ll \Delta_{P}$ \tabularnewline
\hline 
 $g < \frac{1}{M}$ &  $\frac{1}{M} \!  < \!  g \! < \!  1 \! - \! (\frac{\Delta_P}{E_C})^{\frac{2}{M}}$ & $g < \frac{E_C}{M \Delta_P}$  \tabularnewline
\hline 
 $\lambda \! =  \! \frac{g \Delta_P}{E_{C}}$, Sec.~\ref{subsec:WeakSCCoulomb-valley} & $\lambda \! = \! \frac{ \Delta_P}{M \Gamma}$, Sec.~\ref{subsec:2Strong-tunneling-limit} & $\lambda \! = \! \frac{g \Delta_P}{E_{C}}$, Sec.~\ref{subsec:1Weak-tunneling-limit} \tabularnewline
\hline 
 \end{tabular} 
 \caption{
 The cotunneling amplitude $\lambda(D_0)$ in Eq.~(\ref{eq:SummaryTk}) in the conductance intervals  where the Kondo regime exists, see Fig.~\ref{fig:gdiagram}. 
 Here $\Gamma \sim E_C (1-g)^{M/2}$ is the renormalized charging energy in the case $E_C \gg \Delta_P$, see Eq.~(\ref{eq:Gamma}). 
    } \label{table}
\end{table}

\subsection{Average charge \label{sec:SummaryAvN}}

\begin{figure}[tb]
\includegraphics[width=1.0\columnwidth]{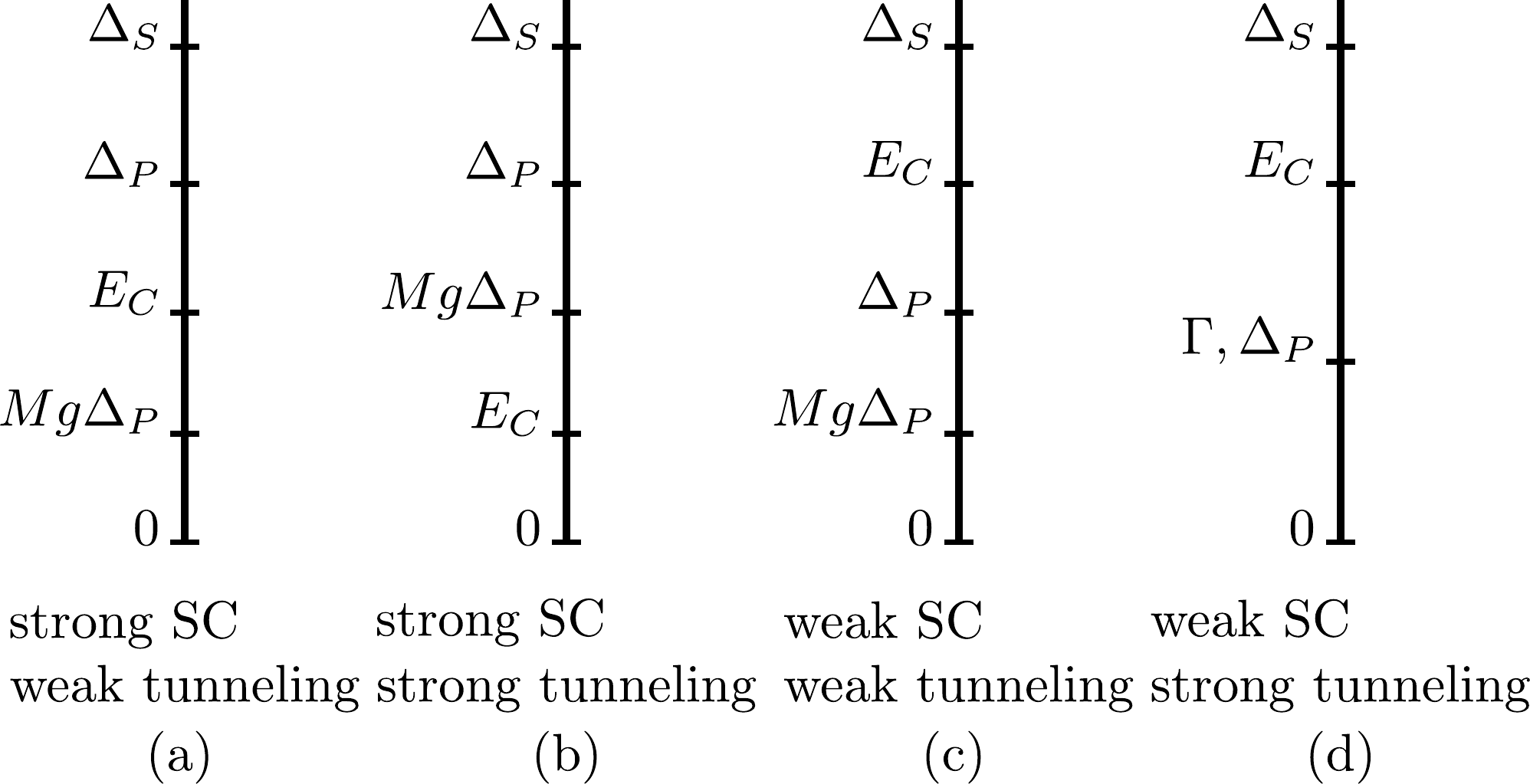}
\caption{The different regimes of parameters that we focus on. These regimes (a)--(d) are discussed in respective Sections~\ref{subsec:1Weak-tunneling-limit}, \ref{subsec:1Strong-tunneling-limit}, \ref{subsec:2Weak-tunneling-limit}, and \ref{subsec:2Strong-tunneling-limit}. 
In the case (d) (weak superconductor, strong tunneling) we have two subcases, $\Gamma = E_C (1-g)^{M/2}$ can be much smaller or much larger than $\Delta_P$.  
\label{fig:regimes}}
\end{figure}
Let us first discuss the average number of electrons $ \langle N \rangle$ of the island, whose dependence on the gate charge $N_g$ can be measured by charge sensing~\cite{PhysRevApplied.11.064011}. 
We focus on the regime  $T \ll E_C$ where thermal excitations can be neglected. 
In the limit $\Delta_P \gg E_C$, the charge depends on the ratio $E_C/\Delta_P$. 
For example, in the strong tunneling limit $g\Delta_P \gg E_C$ [case (b) in Fig.~\ref{fig:regimes}], we have   for all $N_g$,
\begin{equation}
\left\langle N\right\rangle -N_{g} \propto  - \left(\frac{E_C}{\Delta_P}\right)^{M} r_1 \dots r_M \sin2\pi N_{g} \,, \label{eq:chargeIntro}
\end{equation}
where $r_i \sim (1-g_i)^{1/2}$ are the bare reflection amplitudes of the $M$ junctions. 
Thus, the charge of the island almost linearly follows the gate charge $N_g$, apart from harmonic corrections which are weakened by a factor $(E_C/\Delta_P)^M$ due to the enhanced superconducting charge fluctuations. 
The power-law-dependence on $E_C/\Delta_P$ is a signature of topological ground state degeneracy and can be measured in charge-sensing. 
In the opposite limit of weak superconductivity, $E_C \gg \Delta_P$, this factor is absent from Eq.~(\ref{eq:chargeIntro}) [case (d) in Fig.~\ref{fig:regimes}]. 
This result is similar to the one predicted in metallic multi-lead quantum dots~\cite{PhysRevLett.82.1245}. 

In the weak-tunneling limit, $g \Delta_P \ll E_C$, the charge is approximately quantized to the nearest integer to $N_g$ and $\langle N \rangle$ vs $N_g$ shows the usual Coulomb staircase. 
The plateaus in the Coulomb staircase at $N_g \approx \mathrm{integer}$ are not horizontal  
but have a slope $d\left\langle N\right\rangle/dN_g  \approx   \sum_{i} g_i  \Delta_P /E_C$ from quantum fluctuations due to tunneling. 
The zero-temperature slope is thus enhanced by an additional factor $\Delta_P /E_C \gg 1$ compared to the case of a normal metal quantum dot~\cite{glazman1990lifting,PhysRevB.51.1743}. 
In the weak pairing limit ($\Delta_P \ll E_C $) we find indeed $d\left\langle N\right\rangle/dN_g  \approx   \sum_{i} g_i$. 
For more details, see Secs.~\ref{sec:StrongSC_charge}--\ref{sec:weakSC_charge}.

\subsection{Multi-terminal conductance \label{sec:SummaryG}}



One of the signatures of the topological ground-state degeneracy is a universal zero-temperature value of the  off-diagonal elements $G_{i \neq j}$ of the  conductance matrix. (Diagonal elements $G_{ii}$ are determined by current conservation, $G_{ii} = -\sum_{j\neq i} G_{ij}$.) 
Namely, $G_{i \neq j}$ approaches the quantized value $G_{i \neq j}/G_0 = 2/M$ independent of gate voltage $N_g$, $E_C$, $\Delta_P$ and the normal-state conductance $g_i$ of the junctions. 
At finite temperatures, the deviations from the quantized value are 
\begin{equation}
   \frac{G_{i \neq j}}{G_0} - \frac{2}{M}  \propto   - \left( \frac{T} {T_K} \right)^{2(1-\frac{2}{M})} ,\quad (T\ll T_K)  , \label{eq:SummaryGBelowTk}
\end{equation}
where the non-Fermi-liquid type power-law exponent $2(1-\frac{2}{M})$ is a signature of the topological ground-state degeneracy. 
While the dependence on $T/T_K$ is universal, the  temperature $T_K$ depends on $\Delta_P/ E_C$ as shown in Eq.~(\ref{eq:SummaryTk}) and Table~\ref{table} in the Kondo regime. 
In the strong tunneling regime, we have $ T_K \rightarrow \min(E_{C},\,\Delta_{P})$. 
For more details, including the conductance at intermediate temperatures, see Secs.~\ref{sec:ConductanceStrongSC}--\ref{sec:ConductanceWeakSC}.

\section{Theoretical Model \label{sec:model}}

We consider an array of semiconductor nanowires coupled to a conventional s-wave superconductor~\cite{2010PhRvL.105g7001L, Oreg10}. 
The left and right halves of these wires form  $M$ normal metal semi-infinite leads, see Fig.~\ref{fig:device}.
The central region between the leads forms an island which is proximitized by an s-wave superconductor. 
We will focus on energy scales much less than the   Fermi energy, which allows us to linearize the electron spectrum. In this approximation, we can use the bosonization technique~\cite{giamarchi2003quantum}. 

In a bosonized description, the s-wave superconductor can be modeled as having a large spin gap~\cite{2011PhRvB..84s5436F} $\Delta_S$, which we take to be the largest energy scale (UV cutoff) in our system~\footnote{We assume the wires to be farther than the s-wave coherence length from each other so that we can ignore Cooper pair tunneling between them.  }. 
At energies much below $\Delta_S$, only the charge degree of freedom of the superconductor plays a role. 
We can furthermore neglect the spatial dependence of the charge fluctuations by consider a large superconductor with many channels~\cite{2011PhRvB..84s5436F,PhysRevB.97.125404,2019arXiv190910521K}; in this limit the superconductor is fully characterized by its total charge $N_{SC}$ or  its conjugate variable  $\theta_{SC}$. 

We assume that each nanowire is in the helical regime~\cite{2010PhRvL.105g7001L,2010PhRvL.105q7002O} and can  be modeled as a single channel of spinless electrons. 
Under these assumptions we obtain the following low-energy model valid at energies well below $\Delta_S$ (henceforth we set $k_B=\hbar =1$), 
\begin{equation}
H=\sum_{\alpha=1}^{M/2}(H_{0,\alpha}+H_{P,\alpha}+H_{r,\alpha})+E_{C}(N-N_{g})^{2}\label{eq:Hfull} \,,
\end{equation}
	where  $\alpha$ labels the wires and 
	\begin{flalign}
	H_{0,\alpha} & =\frac{v}{2\pi}\int_{-\infty}^{\infty}dx\left([\partial_{x}\theta_{\alpha}(x)]^{2}+[\partial_{x}\varphi_{\alpha}(x)]^{2}\right)\\
	H_{P,\alpha} & =-\frac{\Delta_{P}}{2\pi a}\int_{0}^{L}dx\cos2[\theta_{\alpha}(x) - \theta_{SC}] \label{eq:HP}\\
	H_{r,\alpha} & =-Dr_{\alpha L}\cos2\varphi_{\alpha}(0)-Dr_{\alpha R}\cos2\varphi_{\alpha}(L) \,.
	\end{flalign}
Here $H_{0,\alpha}$ is the kinetic energy of wire $\alpha$,  
$H_{P,\alpha}$ describes the  pair-tunneling to the s-wave superconductor~\cite{2011PhRvB..84s5436F}  and favors a formation of   a superconducting topological gap $\Delta_P$ in the proximitized middle section of wire, $0 < x <L$. 
The proximitized segments of length $L$ are separated from the ``leads'' by barriers that are modeled by the backscattering terms $H_{r,\alpha}$ 
($D = v/a \sim  \Delta_S$ is the cutoff, $r_{\alpha L/R}$ is the dimensionless reflection amplitude of the left/right junction of wire $\alpha$). 
The last term in Eq.~(\ref{eq:Hfull}) is the total charging energy of the  central  island formed by the barriers. In it, $N$ is the total number of electrons   
$N=N_{SC}+ \sum_{\alpha}N_{\alpha}$ 
where in the bosonized language  $N_{\alpha}=\frac{1}{\pi}[\varphi_{\alpha}(L)-\varphi_{\alpha}(0)]$ is the charge of the wire segment $\alpha$. 
The dimensionless gate voltage $N_g$ can be tuned to change the favored values of $N$. 
The bosonic fields of the wires satisfy the commutation relations $[\theta_{\alpha}(x'),\partial_{x}\varphi_{\beta}(x)]=i\pi\delta_{\alpha\beta}\delta(x-x')$. 
For the s-wave superconductor we have $[\theta_{SC},N_{SC}]=i $. 
The pair-tunneling term $H_{P,\alpha}$ therefore conserves the total charge $N$.

We will first derive an effective low-energy boundary model from the Hamiltonian~(\ref{eq:Hfull}). 
For this, it is convenient to move to the   imaginary time action formalism. 
The action obtained from Eq.~(\ref{eq:Hfull}) reads 
	\begin{equation}
	S=\sum_{\alpha=1}^{M/2}S_{\alpha}+\int d\tau E_{C}(N-N_{g})^{2}\,, \label{eq:SplusEc}
	\end{equation}
	where 
	\begin{flalign}
	&S_{\alpha} \! = \! \frac{1}{2\pi}\int \! d\tau \! \! \int_{-\infty}^{\infty} \! \!\! \! dx  \left(v\!\left[[\partial_{x}\theta_{\alpha}]^{2}+[\partial_{x}\varphi_{\alpha}]^{2}\right]+2i\partial_{x}\varphi_{\alpha}\partial_{\tau}\theta_{\alpha}\right)\\
	& - \frac{\Delta_P}{2\pi a}\int d\tau\int_{0}^{L}dx\cos2[\theta_{\alpha}(x) - \theta_{SC}]\\
	& -\! \int d\tau\left[Dr_{\alpha L}\cos2\varphi_{\alpha}(0)+Dr_{\alpha R}\cos2\varphi_{\alpha}(L)\right] , \label{eq:SintroBarrier}
	\end{flalign}
	and throughout all the $\tau$-integrals range from $1/D$ to the inverse temperature $\beta = 1/T$.  
Crucially, the pairing operator $\cos2[\theta_{\alpha}(x) - \theta_{SC}]$ has scaling dimension 1 and as a bulk perturbation is a relevant operator~\cite{2011PhRvB..84s5436F,PhysRevB.97.125404}. 
It reaches strong coupling at bandwidth $D \sim \Delta_P$. For boundary perturbations [e.g. Eq.~(\ref{eq:SintroBarrier})] the marginal dimension is 1. 
At energy scales much above $\Delta_P$ and $E_C$, the barrier term, Eq.~(\ref{eq:SintroBarrier}), as well as the charging energy, Eq.~(\ref{eq:SplusEc}), are both marginal operators with scaling dimension 1.

	We integrate out the lead modes ($x<0$, $x>L$)
	from the first term in $S_{\alpha}$. This leads to the action 
	\begin{equation}
	\bar{S}=\sum_{\alpha=1}^{M/2}\bar{S}_{\alpha}+\int d\tau E_{C}(N-N_{g})^{2}\,, \label{eq:SfullInterm}
	\end{equation}
	where now the total charge of the island 
 	\begin{equation}
 	N=N_{SC}+\frac{1}{\pi}\sum_{\alpha\sigma}\varphi_{\alpha\sigma}\,,
	\end{equation}
	is expressed in terms of the boundary fields 
	\begin{equation}
	\varphi_{\alpha L}=-\varphi_{\alpha}(0)\,,\quad\varphi_{\alpha R}=\varphi_{\alpha}(L)\,,
	\end{equation}
	and the wire action consists of a boundary part and the proximitized middle segment~\footnote{The Fourier transform is defined as \unexpanded{$\varphi_{\alpha\sigma}(\tau)=T\sum_{\omega_{n}}e^{i\omega_{n}\tau}\varphi_{\alpha\sigma}(\omega_{n})$}.}, 
\begin{flalign}
& \bar{S}_{\alpha}=\frac{1}{2\pi}T\sum_{\omega_{n}}\sum_{\sigma=L,R}e^{-|\omega_n|/D} |\omega_{n}||\varphi_{\alpha\sigma}(\omega_{n})|^{2} \label{eq:Seff0}\\
& -\int d\tau\sum_{\sigma=L,R}Dr_{\alpha\sigma}\cos2\varphi_{\alpha\sigma}  \\
& +\int d\tau \int_{0}^{L}  dx\left[\frac{v}{2\pi}\left([\partial_{x}\theta_{\alpha}]^{2}+[\partial_{x}\varphi_{\alpha}]^{2}\right)\right. \label{eq:Seff3}\\
& +\left.\frac{1}{\pi}i\partial_{x}\varphi_{\alpha}\partial_{\tau}\theta_{\alpha}- \frac{\Delta_P}{2\pi a} \cos2[\theta_{\alpha}(x) - \theta_{SC}] \right] \,. \label{eq:Seff4}
\end{flalign}
Here in Eq.~(\ref{eq:Seff0}) we introduced  the bosonic Matsubara frequency $\omega_n = 2\pi n T$ and imposed the cutoff with the  the factor  $e^{-|\omega_n|/D}$. 
Equation~(\ref{eq:Seff0}) is the familiar dissipative action that arises after integrating out the gapless bulk modes~\cite{altland2010condensed}. 
However, the dissipation strength [the prefactor on the first line~(\ref{eq:Seff0})] is half of the usual one for a spinless Luttinger liquid~\cite{PhysRevB.46.15233}. This is due to the fact that we  integrated out modes only on one side of the barrier. 
The next step is to integrate out the proximitized sections $0<x<L$.  
This is in general a difficult task due to the non-linear cosine term in Eq.~(\ref{eq:Seff4}). However, we can resort to approximations at energies much larger or smaller than $\Delta_P$. 
At high energies, $\omega  \gg \Delta_P$, we can ignore the pairing term and the proximitized section contributes to the dissipation due to above-gap quasiparticles. 
This will modify the coefficient in Eq.~(\ref{eq:Seff0}) and is discussed in more detail in Sec.~\ref{sec:Weak-superconductor-limit}. 

At low frequencies, $\omega  \ll\Delta_{P}$, the proximitized segment becomes non-dissipative. 
Indeed, the superconducting pairing makes $\theta_{\alpha}(x)$ uniform throughout the proximitized section of the wire and the action of the $\varphi$-modes in that region becomes non-dissipative, see Sec.~\ref{sec:More-detailed-model} for details. 
We will start by discussing the case of large topological gap, $\Delta_P \gg E_C$, where the physics is governed by this latter limit. 

\section{Strong superconductor limit $\Delta_{P}\gg E_{C}$\label{sec:Strong-superconductor-limit}}
In this Section we focus on the case of large topological gap in comparison to the charging energy, $\Delta_{P}\gg E_{C}$. 
At energy scales $\Delta_P \ll \omega  \ll \Delta_S$ the  nanowires behave essentially as a independent wires in normal state, and $\Delta_P$ can be treated perturbatively~\cite{Note1}. 
At frequencies below the topological gap, $\omega \ll \Delta_P$,  we can ignore above-gap quasiparticle excitations in the island. 
In this case, the proximitized segment, Eqs.~(\ref{eq:Seff3})-(\ref{eq:Seff4}), does not contribute to dissipation in the boundary action, 
see Sec.~\ref{sec:More-detailed-model}. 
We can therefore write a boundary effective action for wire $\alpha$,  
\begin{flalign}
& \bar{S}_{\alpha}=\frac{1}{2\pi}T\sum_{\omega_{n}}\sum_{\sigma=L,R} e^{-|\omega_n|/D} |\omega_{n}||\varphi_{\alpha\sigma}(\omega_{n})|^{2} \label{eq:Sfull0}\\
& -\int d\tau\sum_{\sigma=L,R}Dr_{\alpha\sigma}\cos2\varphi_{\alpha\sigma} \nonumber \,, 
\end{flalign}
and the full action is given by Eq.~(\ref{eq:SfullInterm}) which includes the charging energy that couples the different boundary fields $\varphi_{\alpha\sigma}$.  
The bandwidth $D$ in $\bar{S}_\alpha$ above is now assumed to be well below $\Delta_P$. 

In order to describe energy scales below $E_{C}$, it is convenient
to change variables to a common mode $Q = \sum_j \varphi_j/\sqrt{M}$ and $M-1$ differential
modes, $q_{j}$. 
We denote $j=(\alpha,\sigma)$
which we enumerate as $j=1,\dots,M$ from hereon, see Fig.~\ref{fig:device}. 
The transformation is orthogonal and can be conveniently
written as 
\begin{equation}
\varphi_{j}=\frac{1}{2}\boldsymbol{R}_{j}\cdot\boldsymbol{q}+\frac{1}{\sqrt{M}}Q \,, \label{eq:phiqbasis}
\end{equation}
where $\boldsymbol{q}$ is a $(M-1)$-component vector field and the $M$
vectors $\boldsymbol{R}_{j}$ satisfy $\sum_{j}\boldsymbol{R}_{j}=0$
and $\sum_{j}\frac{1}{4}R_{j}^{a}R_{j}^{b}=\delta^{ab}$ and $\frac{1}{4}{\boldsymbol{R}_{j}\cdot\boldsymbol{R}_{k}}=\delta_{jk}-\frac{1}{M}$.
In terms of these fields the full boundary action obtained from Eq.~(\ref{eq:SfullInterm}) is 
\begin{flalign}
S & =\frac{1}{2\pi}T\sum_{\omega_{n}} e^{-|\omega_n|/D} |\omega_{n}|\left(|\boldsymbol{q}(\omega_{n})|^{2}+|Q(\omega_{n})|^{2}\right)\label{eq:Sfull}\\
& +E_{C}\int d\tau(\frac{1}{\pi}\sqrt{M}Q-N_{g})^{2}\nonumber \\
& -D\int d\tau\sum_{j}r_{j}\cos(\mathbf{R}_{j}\cdot\boldsymbol{q}+2\frac{1}{\sqrt{M}}Q)\,.\nonumber 
\end{flalign}
This action is suitable for perturbative expansion in the reflection amplitudes $r_i \ll 1$, i.e., in the strong tunneling limit. 
For the opposite case of weak tunneling, we introduce a dual action.

In the weak tunneling limit the barriers are high, $r_j \approx 1$ in Eq.~(\ref{eq:Sfull}). In this case it is convenient to use a dual description,
\begin{flalign}
 & S= \frac{T}{2\pi}\sum_{\omega_{n}} e^{-|\omega_n|/D}  |\omega_{n}|\left[ |\boldsymbol{p}(\omega_{n})|^{2} \! + \! \frac{|\omega_{n}| |P(\omega_{n})|^{2} }{|\omega_{n}|+2\frac{M}{\pi}E_{C}}\right] \! + \! S_W \nonumber \\
& \! - \! \int \! d\tau \sum_{j}D t_{j}\cos\left(\frac{1}{2}\boldsymbol{R}_{j}\cdot\boldsymbol{p}+\frac{1}{\sqrt{M}}P+2\pi WT\tau\right)\!  , \label{eq:Sdual}
\end{flalign}
where again $j=(\alpha,\sigma)$ and  $t_j$ is the dimensionless tunneling amplitude of an electron to pass from lead $j$ to the island. 
The dual fields $\boldsymbol{p},\,P$ are defined analogously to Eq.~(\ref{eq:phiqbasis}), 
\begin{equation}
    \frac{1}{2}\boldsymbol{R}_{j}\cdot\boldsymbol{p}+\frac{1}{\sqrt{M}}P=\delta\theta_{j}\,,
\end{equation}
where we denote $\delta\theta_{\alpha L}=-\theta_{\alpha}(x)|_{0-}^{0+}$  and $\delta\theta_{\alpha R}=\theta_{\alpha}(x)|_{L-}^{L+}$. 
Then $[\varphi_{j},\delta\theta_{k}]=i\pi\delta_{jk}$ and as a result, $[q_j, p_k]=i\pi \delta_{jk}$ and  $[Q,P]=i\pi$. 

The action~(\ref{eq:Sdual}) allows us to use perturbation theory in the weak tunneling limit, where $t_i \ll 1$.   
In Eq.~(\ref{eq:Sdual}) charge quantization is imposed by the  integer winding number $W$ whose   action contains the dependence on gate charge $N_g$:   $S_W = -iN_{g}2\pi W + \pi^{2}TE_{C}^{-1}W^{2}$. 
The partition function includes a sum over winding numbers,  $Z  =\sum_{W=-\infty}^{\infty}\int \mathcal{D} \boldsymbol{p} \mathcal{D} P e^{-S}$. 
The dual action is derived by performing a Villain
transformation in Eq.~(\ref{eq:Sfull}), see Appendix~\ref{sec:Detailed-derivation-of} for details.

The discussion in the next sections is entirely based on analyzing the actions~(\ref{eq:Sfull}) and~(\ref{eq:Sdual}) by using  perturbative renormalization group.

\subsection{Weak tunneling limit $Mg \Delta_{P} \ll E_{C} $ \label{subsec:1Weak-tunneling-limit}}

The weak tunneling regime [$r_{j}\approx1$ in Eq.~(\ref{eq:Sfull})]  is the previously studied conventional topological Kondo limit~\cite{2012PhRvL.109o6803B,2013PhRvL.110u6803B,2013PhRvL.110s6401A}. 
It is thus convenient to use the dual action Eq.~(\ref{eq:Sdual}). 
Alternatively one can use a fermionic description~\cite{2016arXiv160800581M},
see Eq.~(\ref{eq:StFermion}) below.

At high energies $\omega_n \gg E_{C}$, the tunneling operator in the second line of Eq.~(\ref{eq:Sdual}) has scaling dimension $\Delta=\frac{1}{2}\frac{1}{4}|\mathbf{R}_{i}|^{2}+\frac{1}{2}\frac{1}{M}=\frac{1}{2}$ so it is a RG relevant boundary perturbation. 
This is due to the superconducting
gap $\Delta_{P}$ that pins the field in the proximitized segment
of the wire. 
From the RG equation, 
\begin{equation}
    \frac{dt_i}{dl} = \frac{1}{2} t_i\,, \quad(l=-\ln D/\Delta_{P})\,, \label{eq:tiRunning}
\end{equation}
we obtain the running coupling  $t_{i}(D) \sim \sqrt{g_{i}}\sqrt{\Delta_{P}/D}$ in terms of the running cutoff $D$. 
We set here the bare value, $t_{i}(\Delta_{P})^{2}= g_{i}/\pi^2$, to be given by  the normal state dimensionless conductance $g_{i}$ of the junction. 
The scaling dimension $1/2$ corresponds to tunneling into a Majorana state: indeed in the
fermion language we have~\cite{2016arXiv160800581M} 
\begin{equation}
S_{t}=-i\int d\tau\sum_{i}Dt_{i}(\psi_{i}N^{+}+\psi_{i}^{\dagger}N^{-})\gamma_{i}\,,\label{eq:StFermion}
\end{equation} 
where $\gamma_{i}$ is a Majorana operator and $\psi_{i}$ is a spinless lead fermion annihilation operator; $N^{+}=(N^{-})^{\dagger}$ raises the total charge of the island by 1.

Let us next consider lower energies with a new cutoff $D\ll E_{C}$ obtained by integrating out high-energy modes.  We   integrate out the field $P$ (which is conjugate to total charge $Q$) in Eq.~(\ref{eq:Sdual}). We find to second order in the tunneling action (see Sec.~\ref{sec:TunAction} for details), 
\begin{widetext}
\begin{flalign}
S_{t,\text{eff}} & =-\frac{1}{4} D^{2}\int d\tau\int d\tau'C(\tau-\tau')\sum_{ij}t_{i}(D)t_{j}(D) 
\cos \left(\frac{1}{2}\boldsymbol{R}_{i}\cdot\boldsymbol{p}(\tau)-\frac{1}{2}\boldsymbol{R}_{j}\cdot\boldsymbol{p}(\tau')\right)
,\label{eq:StEffStrongSC}
\end{flalign} \end{widetext}
where the correlation function 
\begin{equation}
C(\tau)=\left\langle e^{i\frac{1}{\sqrt{M}}[P(\tau)-P]+2\pi iWT|\tau|}\right\rangle _{P,W}\approx   \frac{e^{-E_{C}^{*}|\tau|}}{\sqrt{1+D^2 \tau^{2}}^{1/M}}  , \label{eq:Ctau}
\end{equation}
is valid for $T \ll \tau^{-1},E_C,D$ and is derived in Sec.~\ref{subsec:The-correlation-function}. 
We have introduced $E_{C}^{*}=2(\frac{1}{2}-N_{g})E_{C}$ assuming ${0\leq N_{g}\leq1/2}$. 
The correlation function $C(\tau)$ is factored into algebraically and exponentially decaying parts. The former comes from the dissipative dynamics of the field $P$ while the latter comes from the dynamics of its charge. 

Depending on $E_{C}^{*}$, ultimately determined by $N_{g}$, the weak tunneling limit can be
divided into two regimes: 1. the case of \emph{Coulomb valley} where
$N_{g}$ is close to an integer and the charge of the island cannot
change without a large energy cost $E_C$, and 2. the case of \emph{charge
degeneracy} where $N_{g}$ is close to a half-integer and the island
charge can fluctuate between two values. We will start our discussion
from the first regime of Coulomb valley. Note that even in the Coulomb
valley, the island still has ground state degeneracy due to the many
Majorana zero modes. This ground state degeneracy gives rise to a
topological Kondo effect, distinct from the charge Kondo effect. 

\subsubsection{Coulomb valley\label{subsec:StrongSCCoulomb-valley}}

When $D\ll E_{C}^{*}$, we have $|\tau-\tau'| \lesssim  E_{C}^{*-1} \ll D^{-1}$
in Eqs.~(\ref{eq:StEffStrongSC})--(\ref{eq:Ctau}). 
The correlation function can then be approximated as $C(\tau) \approx e^{-E_{C}^{*}|\tau|}$. 
Due to its fast exponential decay, we can set $\tau\approx\tau'$ in the $\boldsymbol{p}$-fields in Eq.~(\ref{eq:StEffStrongSC}). 
We then find 
\begin{equation}
S_{t,\text{eff}}=-\frac{1}{4}D\int d\tau\sum_{ij}\lambda_{ij}(D)\cos\frac{1}{2}[\boldsymbol{R}_{i}-\boldsymbol{R}_{j}]\cdot\boldsymbol{p}(\tau)\,,\label{eq:SStrongSCvalley}
\end{equation}
where $\lambda_{ij}$ has a bare value  $\lambda_{ij}(E_C^*) \approx \sqrt{g_i g_j} \Delta_P /E_C^*$; we  used $\int d\tau C(\tau)= 1/E_{C}^{*}$. 
We can analyse Eq.~(\ref{eq:SStrongSCvalley}) perturbatively as
long as $\lambda_{ij}(D)\ll1$. The operator in Eq.~(\ref{eq:SStrongSCvalley})
is marginal as $\Delta=\frac{1}{2}\frac{1}{4}|\boldsymbol{R}_{i}-\boldsymbol{R}_{j}|^{2}=1$. 
It turns out to be marginally relevant~\cite{anderson_poor_1970,2016PhRvB..94w5102H},
with the RG equation 
\begin{equation}
\frac{d}{dl}\lambda_{ij}=2\sum_{k\neq i,j}\lambda_{ik}\lambda_{kj}\,.\label{eq:topoKondoRG}
\end{equation}
The system flows to a new fixed point of large $\lambda_{ij}$ where the electron number parity
on the island is strongly coupled with the parity in the leads. This
phenomenon is called the topological Kondo effect~\cite{2012PhRvL.109o6803B}.
Along the flow to the new fixed point, the couplings flow towards
isotropic line $\lambda_{ij}\to\lambda$ and the flow equation~(\ref{eq:topoKondoRG}) becomes
\begin{equation}
\frac{d\lambda}{dl}=2(M-2)\lambda^{2}\,,\label{eq:topoKondoRGIso}
\end{equation}
 with solution 
 $\lambda(D)=\lambda(E_C^*) \ln (e E_C^*/T_K )/\ln (e D/T_K) $. 
 We define the Kondo temperature as the strong coupling scale at which $\lambda$ becomes large (i.e., $2(M-2)\lambda (T_K) \sim 1$),
\begin{equation}
T_{K}\sim E_{C}^{*}\exp\left[-\frac{1}{2(M-2) g\Delta_{P}/E_{C}^{*}}\right]\,. \label{eq:TKStrongSC}
\end{equation}
 We expressed $T_{K}$ in terms of the bare conductance by using
$t_{i}(D)=\sqrt{g_{i}}\sqrt{\Delta_{P}/D}$ and taking approximately
isotropic barriers. In Eq.~(\ref{eq:topoKondoRGIso}) we assume for
the bare coupling that $\lambda(E_{C}^{*})\ll1/2(M-2)$.  
This defines the Coulomb valley regime where perturbative
expansion of Eq.~(\ref{eq:SStrongSCvalley}) is valid, $Mg\Delta_{P}\ll E_{C}^{*}$. 
 Therefore we can use this to estimate the width $|\frac{1}{2}-N_{g}^{*}|$
in gate voltage of the regime of strong charge fluctuations around
the charge degeneracy point,
\begin{equation}
\left|\frac{1}{2}-N_{g}^{*}\right|\sim Mg\Delta_{P}/E_{C}\,.\label{eq:peaksize}
\end{equation}
 The weak tunneling limit is defined by the condition $Mg\ll E_{C}/\Delta_{P}$
which also implies $T_{K}<Mg\Delta_{P}$. 

The energy scale $Mg\Delta_{P}$ that enters in Eq.~(\ref{eq:peaksize})
also appears in observable quantities (Sec.~\ref{subsec:StrongSCAverage-charge-and})
and upon comparing to $E_{C}$ distinguishes the weak and strong tunneling limits. 
The scale can be interpreted as the broadening of the ground state manifold due to  Majorana-lead couplings.  
Our Hamiltonian~(\ref{eq:Hfull}) allowed us to connect this scale to the experimentally accessible
microscopic parameters $\Delta_{P}$ and $g$; this connection is
beyond the earlier works~\cite{2012PhRvL.109o6803B,2013PhRvL.110u6803B,2013PhRvL.110s6401A,2014PhRvL.113g6401A,2016PhRvB..94w5102H,2016arXiv160800581M}
where the superconducting gap does not enter since the limit $\Delta_{P}\to\infty$
is assumed from the start. 

Near the charge degeneracy point, $N_{g}^{*}<N_{g}<1/2$, we cannot
use the perturbative Eq.~(\ref{eq:SStrongSCvalley}). 
This limit is studied in the next section.

\subsubsection{Charge degeneracy point \label{subsec:StrongScCD}}

Near $N_{g}=1/2$, the island charge is allowed to fluctuate between
two values $0$ and $1$, and $E_{C}^{*}\ll T$ in Eq.~(\ref{eq:StEffStrongSC})
is a small energy scale. We have then $\tau E_{C}^{*}\ll1$ and the
correlation function $C(\tau)\approx(D|\tau|)^{-1/M}$ decays slowly
due to the dissipative dynamics of the field $P$; one then cannot set $\boldsymbol{p}(\tau)$ and $\boldsymbol{p}(\tau')$ equal in Eq.~(\ref{eq:StEffStrongSC}). 
The tunneling
action $S_{t}$ is then most convenient to present in  the fermionic
form given in Eq.~(\ref{eq:StFermion}) upon projecting it to two
charge states, $N^{+}\to\sigma^{+}$, where the operator  $S^{+} =S_x + i S_y =|1\rangle\langle0|$
acts on the subspace spanned by the charge states $0$ and $1$. 
Close to the degeneracy point the effective Hamiltonian can be written as
\begin{equation}
    H_{CK} = \sum_{i} D \left[ J_{\perp i}\frac{1}{2}(s_{i}^{-}S^{+}+s_{i}^{+} S^{-}) + J_z s_{z,i} S_z \right] +  \gamma S_z ,  \label{eq:StrongSCHCK}
\end{equation}
where we introduced the pseudospin operators $s_i^- = (s_i^+)^\dagger = (-i) \psi_{i} \gamma_i $ and $s_{z,i} = \frac{1}{2} - \psi^{\dagger}_{i} \psi_{i}   $. 
The first term, $J_{\perp i} \propto t_i$, in Eq.~(\ref{eq:StrongSCHCK}) arises from Eq.~(\ref{eq:StFermion}) while the second term is zero in the bare Hamiltonian, $J_{z}=0$, but will be generated in the RG procedure. 
The last term in Eq.~(\ref{eq:StrongSCHCK}) is the projected charging energy and 
takes the form of an  effective Zeeman energy caused by small non-resonant gate voltage $\frac{1}{2}-N_{g}$.

Assuming approximately isotropic tunnel couplings in Eq.~(\ref{eq:StrongSCHCK}), one arrives at the following RG equations~\cite{2016arXiv160800581M,2016PhRvB..94w5102H}
\begin{flalign}
\frac{dJ_{\perp}}{dl} & =\frac{1}{2}J_{\perp}+J_{\perp}J_{z}(1-\frac{M}{2}J_{z})  - \frac{M}{2} J_{\perp}^3\,,\label{eq:StrongSCPeakRG}\\
\frac{dJ_{z}}{dl} & =J_{\perp}^{2}(1-MJ_{z})\,, \label{eq:StrongSCPeakRG2} \\ 
\frac{d\gamma}{dl} & =\gamma(1-\frac{1}{2}MJ_{\perp}^{2})\,, \label{eq:gammaRG}
\end{flalign}
where $l=-\ln D/E_{C}$ and we have the initial conditions $J_{\perp}(0) \sim \sqrt{g_i \Delta_P/E_C}$, $J_z(0)=0$, and $\gamma(0) = E_C^*/E_C$. 

Let us analyze the RG Equations~\eqref{eq:StrongSCPeakRG}-- \eqref{eq:StrongSCPeakRG2} in the limit when $M \gg 1$\footnote{one may notice some similarity of Eqs.~\eqref{eq:StrongSCPeakRG}-- \eqref{eq:StrongSCPeakRG2} with the charge Kondo RG equations~\cite{matveev1991quantum,2002PhRvB..65s5101Y}, see Eqs.~(\ref{eq:WeakSCPeakRG})--(\ref{eq:WeakSCPeakRGgamma}).  An important effect of topological superconductivity is that 
the coupling $J_{\perp}$ now acquires trivial scaling dimension $1/2$ which significantly modifies the RG flow.}. It is useful to rescale exchange couplings $\tilde{J}_\perp \rightarrow \sqrt{M}J_\perp$ and $\tilde{J}_z \rightarrow M J_z$ to find
\begin{flalign}
\frac{d \tilde{J}_{\perp}}{dl} & =\frac{1}{2}\tilde{J}_{\perp}+\frac{1}{M}\tilde{J}_{\perp}\tilde{J}_{z}(1-\frac{1}{2}\tilde{J}_{z})  - \frac{1}{2} \tilde{J}_{\perp}^3\,,\label{eq:StrongSCPeakRG11}\\
\frac{d\tilde{J}_{z}}{dl} & =\tilde{J}_{\perp}^{2}(1-\tilde{J}_{z})\,, \label{eq:StrongSCPeakRG22}
\end{flalign}
The cross terms in Eq.~\eqref{eq:StrongSCPeakRG11} are $O(1/M)$ and can be neglected yielding a closed equation for $\tilde{J}_{\perp}(l)$: 
\begin{align}\label{eq:solutionTK}
\left . \left( \ln x-\frac{x^2}{2}\right)\right|_{x = \tilde{J}_{\perp}(0)}^{x = \tilde{J}_{\perp}(l)}=\frac{l}{2} \, ,    
\end{align}
where the initial condition is $\tilde{J}_{\perp}(0)\sim \sqrt{M g \Delta_P/E_C}$. Assuming $\tilde{J}_{\perp}(0) \ll 1$, the solution for $\tilde{J}_{\perp}(l)$ reads
\begin{align}\label{eq:sol}
\tilde{J}_{\perp}(l)\approx \left \{ \begin{array}{cc}
                 \sqrt{M g \Delta_P/E_C} e^{l/2}, & \tilde{J}_{\perp}(l) \ll 1 \\
                 1-\frac{\sqrt{\ln \left(E_C / M g \Delta_P\right)-l}}{\sqrt 2}, & |1-\tilde{J}_{\perp}(l)|\ll 1 
               \end{array}
 \right.
\end{align}
The Kondo temperature  can be obtained by   solving $\tilde{J}_{\perp}(T_K)\approx 1$, yielding 
\begin{align}
T_K\sim Mg\Delta_P.     \label{eq:TKPeakStrongSC} 
\end{align}
This estimate for the Kondo temperature at the charge degeneracy point  matches with the Kondo temperature in the valley, Eq.~(\ref{eq:TKStrongSC}),  at $E_{C}^{*}\sim Mg\Delta_{P}$, which suggests a width $\sim \! Mg\Delta_{P}/E_C$ of the charge degeneracy region.

We can further verify this estimate by studying the Zeeman term~(\ref{eq:WeakSCPeakRGgamma}). 
By using the solution \eqref{eq:sol},
we find $\gamma(l)=\gamma(0)\exp\left(l-\frac{1}{2}[e^{l}-1]\tilde{J}_{\perp}(0)^{2}\right)$. 
This shows that initially $\gamma$ grows (since it is a relevant perturbation) but
its flow reverses due to quantum fluctuations described by the fast-growing
$J_{\perp}^{2}$ -term. By using $J_{\perp}(0)^{2}=g\Delta_{P}/E_{C}$
we find that the maximum value $\gamma\sim E_{C}^{*}/Mg\Delta_{P}$
is reached at scale $Mg\Delta_{P}$ {[}which is the Kondo temperature at charge degeneracy, see Eq.~(\ref{eq:TKPeakStrongSC}), and where $J_{\perp}$ becomes of order $1/\sqrt{M}${]}. 
When $E_{C}^{*}\ll Mg\Delta_{P}$, the
quantum fluctuations are strong enough so that $\gamma$ never reaches
a large value. This gives a cross over scale $E_{C}^{*}\sim Mg\Delta_{P}$
between the Coulomb valley and charge degeneracy, which agrees with
our estimate~(\ref{eq:peaksize}) obtained from the RG equations
in the valley.

When the bare conductance is increased to of order $g\sim E_{C}/(M\Delta_{P})$,
the region of strong charge fluctuations covers the entire Coulomb
valley and the valley Kondo temperature becomes of order $T_{K}\sim E_{C}$. 
The perturbative Eq.~(\ref{eq:StrongSCPeakRG}) is no longer valid as $J_\perp (0) \sim \sqrt{1/M}$.  
This conductance scale is the cross over to strong tunneling, studied in Sec.~\ref{subsec:1Strong-tunneling-limit}.

 In the strong coupling limit, $T \ll T_K$ when  $J_{\perp}\sim1/\sqrt{M}$, one can use perturbation theory in weak irrelevant reflection operators which we will discuss in the next section~\ref{sec:StrongCoupling}.   
 This is conceptually very similar to the discussion in Sec.~\ref{subsec:1Strong-tunneling-limit} as well.

\subsubsection{The strong-coupling limit \label{sec:StrongCoupling}}

As shown in the previous section at low energies the system flows to a strong coupling fixed point which is characterized by the presence of a Kondo resonance regardless of the gate charge $N_g$. The leading irrelevant perturbations around the fixed point correspond to quasiparticle reflection from the contacts, which is described by the operator $\sum_{j}r_{j}\cos(\mathbf{R}_{j}\cdot\boldsymbol{q}+2\frac{1}{M}\pi N_{g})$ of scaling dimension $\Delta = 2(1- \frac{1}{M})$, see Sec.~\ref{subsec:1Strong-tunneling-limit} below. 
The $N_g$-dependence can be interpreted as a Berry phase~\cite{2016PhRvB..94w5102H} and it changes the QBM lattice from triangular to honeycomb in the charge degeneracy point. However, $N_g$ only plays a role in $M$th order perturbation theory, see Sec.~\ref{subsec:StrongSCAverage-charge-and}. 
We can estimate the ``bare'' amplitude $r_i (T_K)$ by assuming that the dual description matches with the tunneling formulation at scale $T_K$, i.e.,  $r_i (T_K) \sim 1 $ (assuming $t(T_K) \sim 1/\sqrt{M} $). 
We thus obtain the running coupling  $r_i (D) \sim  (D/T_K)^{1- \frac{2}{M}}$, which will determine the temperature-dependence of observables, see Sec.~\ref{subsec:StrongSCAverage-charge-and}. 

The dual field scaling dimension $\Delta_{r} = 2(1-\frac{1}{M})$ obtained above satisfies the relation~\cite{2002PhRvB..65s5101Y} 
\begin{equation}
\Delta_{r}\Delta_{t}=\begin{cases}
(1-\frac{1}{M})^{2}\,, & \text{(charge degeneracy)}\\
2(1-\frac{1}{M})\,, & \text{(valley)} 
\end{cases}\label{eq:Yi}
\end{equation}
where $\Delta_{t}$ is the scaling dimension of the tunneling amplitude, equaling 1 in the valley, Sec.~\ref{subsec:StrongSCCoulomb-valley}, and $\frac{1}{2}(1-\frac{1}{M})$ at charge degeneracy,  Sec.~\ref{subsec:StrongScCD}.  
This  is a geometric relation for quantum Brownian motion in a hyperhoneycomb
(charge degeneracy) or hypertriangular (valley) lattice, see Ref.~\onlinecite{2002PhRvB..65s5101Y}.

\subsection{Strong tunneling limit $Mg \Delta_{P} \gg E_{C}$\label{subsec:1Strong-tunneling-limit}}

We now consider the case where the quantum fluctuations of charge
on the island are large for all values of $N_{g}$. This corresponds
to bare conductances $g\gg E_{C}/(M\Delta_{P})$. 
In this limit there is  no charge quantization or Coulomb blockade of tunneling since the  broadening of the ground state manifold $Mg\Delta_{P}$ exceeds $E_{C}$. 
The weak-coupling Kondo regime discussed in Sec.~\ref{subsec:StrongSCCoulomb-valley} also does not exist. 
Note that the bare conductance does not have to be large in this limit.  For example if we have $ 1 \gg g \gg E_C / (M \Delta_P)$, we can still use the weak-tunneling action~(\ref{eq:Sdual}). With the RG equation for tunneling amplitude, Eq.~(\ref{eq:tiRunning}), we obtain the strong coupling scale $ g \Delta_P$.  
Below this scale, tunneling is strong and one may use the dual action~(\ref{eq:Sfull}) with the scale  $g \Delta_P$ as the UV cutoff. 

When $M$ is large and $E_C / \Delta_P \gg g \gg  E_C / (M \Delta_P)$, tunneling is seemingly still weak at $D \sim E_C \gg g \Delta_P$ and $r(E_C) \sim 1$. However, the perturbation theory of Sec.~\ref{subsec:StrongSCCoulomb-valley} nevertheless fails due to the large number $\sim M$ of terms. Thus, in this limit it still makes sense to identify $T_K \sim E_C$. 

Let us next consider even higher conductance, $g \gg E_C / \Delta_P$. 
Below the energy scale $g \Delta_P$, we use the dual action 
Eq.~(\ref{eq:Sfull}). 
At energy scales $\omega \gtrsim E_C$ we can ignore the charging energy and find that the scaling dimension of the reflection operator is $\Delta=\frac{1}{2}|\mathbf{R}_{j}|^{2}+\frac{1}{2}4\frac{1}{M}=2>1$  and reflection is irrelevant. 
Thus, the reflection amplitude at scale $E_C$ is $r(E_C) \approx  E_C/(g \Delta_P)$, where we took the UV cutoff to be $ g \Delta_P$ and assumed that $r(g \Delta_P) \sim  1$. This inequality maybe satisfied even when $g \ll 1$. 
In the limit $g \to 1$, we can use $\Delta_P$ as the UV cutoff and a bare value $r(\Delta_P) \approx \sqrt{1-g}$. 
We then have $r(E_C) =(E_C/{\Delta_P} ) \sqrt{1-g} $. The two expressions for $r(E_C)$ match when $g\sim 1$. 


At low energies $\omega \ll E_{C}$ we can ``integrate out'' the total charge mode $Q$~\cite{PhysRevB.51.1743,matveev1994charge}. 
For this, we use the average (see Appendix~\ref{subsec:CorrFnQ}), 
\begin{equation}
\left\langle e^{i2Q/\sqrt{M}}\right\rangle_Q =\left(\frac{2 e^{\gamma}ME_{C}}{\pi D}\right)^{2/M}e^{i2\frac{1}{M}\pi N_{g}}\,, \label{eq:QmassiveSC}
\end{equation}
where $\gamma\approx0.577$ is the Euler\textendash Mascheroni constant and the average is performed with respect  to the action~(\ref{eq:Sfull}). 
The action~(\ref{eq:Sfull})  averaged over $Q$ becomes then
\begin{flalign}
S & =\frac{1}{2\pi}T\sum_{\omega_{n}}e^{-|\omega_n|/D} |\omega_{n}||\boldsymbol{q}(\omega_{n})|^{2} -\frac{D^2}{E_C}\left(\frac{2e^{\gamma}ME_{C}}{\pi D}\right)^{2/M} \nonumber \\
 & \times \negthinspace\int\negthinspace d\tau\sum_{j}r_{j}(E_C)\cos(\mathbf{R}_{j}\cdot\boldsymbol{q}+2\frac{1}{M}\pi N_{g})\,, \label{eq:StrongDeltaStrongTunnelingAction}
\end{flalign}
where $r_i(E_C)$ is the renormalized reflection amplitude described above. 
The dimension of reflection operator is now $\Delta = \frac{1}{2} |\mathbf{R}_{i}|^{2} =2(1-\frac{1}{M})$
because of the pinned field $Q$ by charging energy. 
Despite this reduced scaling dimension, the reflection remains irrelevant. 
Thus,  there is  only weak Coulomb blockade at low energies~\cite{aleiner2002quantum}. 
The weak reflection gives rise to weak harmonic corrections to the
average charge, $\frac{1}{\pi}\sqrt{M}\left\langle Q\right\rangle -N_{g}\propto r_{1}\dots r_{M}\sin2\pi N_{g}$, see Sec.~\ref{subsec:StrongSCAverage-charge-and}.

\section{Weak superconductor limit $\Delta_{P}\ll E_{C}$\label{sec:Weak-superconductor-limit}}

In this Section we focus on the case of a small topological gap in comparison to the charging energy, $\Delta_{P} \ll E_{C}$. 
At energy scales below the s-wave gap and still above charging energy $E_C$, we can use the  boundary action~(\ref{eq:Seff0})--(\ref{eq:Seff4}) and neglect the p-wave pairing term $\propto \Delta_P$. 
After integrating out the modes in the middle segment ($0<x<L$), we obtain an action similar to Eq.~(\ref{eq:Sfull0}) except that the dissipative action will have an additional prefactor 2 due to the dissipative fluctuations from both sides of the barriers.  
By using the procedure outlined in  Sec.~\ref{sec:Strong-superconductor-limit}, we obtain the effective boundary action 
\begin{flalign}
S & =\frac{1}{\pi}T\sum_{\omega_{n}} e^{-|\omega_n|/D} |\omega_{n}|\left(|\boldsymbol{q}(\omega_{n})|^{2}+|Q(\omega_{n})|^{2}\right)\label{eq:SfullWeakSC}\\
& +E_{C}\int d\tau(\frac{1}{\pi}\sqrt{M}Q-N_{g})^{2}\nonumber \\
& -D\int d\tau\sum_{j}r_{j}\cos(\mathbf{R}_{j}\cdot\boldsymbol{q}+2\frac{1}{\sqrt{M}}Q)\,,\nonumber 
\end{flalign}
in the strong tunneling (weak barrier) limit.  
In the weak tunneling regime we find instead [compare to Eq.~(\ref{eq:Sdual})]
\begin{flalign}
 & S= \frac{T}{4\pi}\sum_{\omega_{n}} e^{-|\omega_n|/D}  |\omega_{n}| \! \left[ |\boldsymbol{p}(\omega_{n})|^{2} \! + \! \frac{|\omega_{n}| |P(\omega_{n})|^{2} }{|\omega_{n}|+\frac{M}{\pi}E_{C}}\right] \! + S_W \nonumber \\
& \! - \! \int \! d\tau \sum_{i}D t_{i}\cos\left(\frac{1}{2}\boldsymbol{R}_{i}\cdot\boldsymbol{p}+\frac{1}{\sqrt{M}}P+2\pi WT\tau\right)\!  , \label{eq:SdualWeakSC}
\end{flalign}
which is derived from Eq.~(\ref{eq:SfullWeakSC}) by means of a duality transformation as outlined in Appendix~\ref{sec:Detailed-derivation-of}. 
In Eq.~(\ref{eq:SdualWeakSC}) we have the same winding number action as in  Eq.~(\ref{eq:Sdual}), $S_W = -iN_{g}2\pi W + \pi^{2}TE_{C}^{-1}W^{2}$, and the partition function includes a sum over winding numbers,  $Z  =\sum_{W=-\infty }^{\infty}\int \mathcal{D} \boldsymbol{p} \mathcal{D} P e^{-S}$. 

Equations~(\ref{eq:SfullWeakSC})--(\ref{eq:SdualWeakSC}) are  valid at energy scales much above the topological gap, $\omega \gg \Delta_P$, since we ignored the pairing in them. 
At low temperatures $T \ll \min(\Delta_P,E_C)$ we expect to find the same universal features  in both limits of strong (Sec.~\ref{sec:Strong-superconductor-limit}) and weak superconductor. 
Qualitative differences between the strong and weak superconductors limits are found at temperatures $T \gg \Delta_P$. 
At such temperatures in the weak superconductor limit, the island behaves like a normal state metallic quantum dot. 
The normal state quantum dots have been previously studied by Matveev and Furusaki~\cite{matveev1991quantum,PhysRevB.51.1743,matveev1994charge,1995PhRvB..5216676F}
in the 2-lead case and by Yi and Kane~\cite{1998PhRvB..57.5579Y,2002PhRvB..65s5101Y} in the many-lead case (see also Ref.~\onlinecite{AFFLECK2001535}). 
The weak topological superconducting limit has been studied for two leads by Lutchyn and Glazman~\cite{2017PhRvL.119e7002L}. 
It is particularly interesting that in the temperature-interval $\Delta_P \ll T \ll E_C$ in the Coulomb valley, the junction reflections are  \textit{relevant} perturbations (and tunneling irrelevant) and thus the conductance   decreases upon lowering the temperature. 
However,  upon crossing the scale $\Delta_P$, the situation reverses when reflections become irrelevant and tunneling becomes relevant, as discussed in  Sec.~\ref{sec:Strong-superconductor-limit}. 
This may result in  non-monotonic temperature-dependence of the  conductance as we will see in  Sec.~\ref{sec:ConductanceWeakSC}. 
Furthermore, at charge degeneracy point, the intermediate fixed point of the multichannel Kondo effect~\cite{1998PhRvB..57.5579Y,2002PhRvB..65s5101Y} may be (almost) reached if the corresponding Kondo temperature lies between $\Delta_P$ and $E_C$, see Sec.~\ref{sec:ConductanceWeakSC}.

\subsection{Weak tunneling limit $Mg \ll 1$\label{subsec:2Weak-tunneling-limit}}

We start from the tunneling action~(\ref{eq:SdualWeakSC})  at high frequencies $\omega  \gg \Delta_P$. 
Then, at $\omega \gg E_C$, we can ignore  the charging energy and find the scaling dimension of the tunneling operator  $\Delta=\frac{1}{2}2(\frac{1}{4}|\mathbf{R}_{j}|^{2}+\frac{1}{M})=1$, as expected from the non-interacting limit. 
Thus, the tunneling amplitude does not get renormalized and we have $t_{i}(D)=\sqrt{g_{i}}$ independent of the running cutoff $D$ for  $D\gtrsim E_{C}$. 
Upon integrating out high-energy modes and reducing the cutoff, in the interval $\Delta_{P}\ll D\ll E_{C}$ 
the scaling dimension of the tunneling perturbation increases and Coulomb blockade of conductance may emerge as discussed in detail in the below sections. 

Upon lowering the cutoff scale further, below the topological gap, $D\ll\Delta_{P}$,
the superconducting pairing gaps out the fluctuating modes $\boldsymbol{p}$
in half of the wire, which halves the dissipation strength in the action~(\ref{eq:SdualWeakSC}). Consequently,    the scaling dimensions
of tunneling (reflection) operators are lowered (increased) and 
it  becomes relevant, $\Delta < 1$. 
Likewise we will find (see below) that reflection is irrelevant and there is thus a stable  fixed point without  Coulomb blockade oscillations of conductance, even though we started from weak bare tunnelings. 

Next, we will   analyze in more detail the tunneling action~(\ref{eq:SdualWeakSC}) in the two cases of Coulomb valley ($N_{g}$ close to integer) and charge degeneracy points ($N_{g}$ close to half-integer).

\subsubsection{Coulomb valley \label{subsec:WeakSCCoulomb-valley}}

Similar to Sec.~\ref{subsec:1Weak-tunneling-limit}, in the frequency interval
$\Delta_{P}\ll \omega \ll E_{C}^{*}$ the field $P$ in Eq.~(\ref{eq:SdualWeakSC}) is massive. 
After integrating out $P$, the action for the gapless modes $\boldsymbol{p}$ is {[}compare to  Sec.~\ref{subsec:StrongSCCoulomb-valley}{]}
\begin{equation}
S\!=\! \frac{T}{4\pi}\sum_{i\omega_{n}}|\omega_{n}||\boldsymbol{p}|^{2}-D\int d\tau\sum_{i\neq j}\lambda_{ij}(D)\cos \frac{1}{2}(\boldsymbol{R}_{i}-\boldsymbol{R}_{j})\cdot\boldsymbol{p}, \label{eq:Tunneling_mu}
\end{equation}
where $\lambda_{ij}(E_C^*)=\sqrt{g_i g_j} \approx g$ 
is the (bare) amplitude to tunnel an electron between leads $i$ and $j$ without changing the total charge of the island. 
In the Coulomb valley, $D<E_{C}^{*}$, the dimension of tunneling
operator above is $\Delta=2$~\cite{1995PhRvB..5216676F,1998PhRvB..57.5579Y} 
and tunneling is  therefore irrelevant, $\lambda_{ij}(D)=gD/E_{C}^{*}$.  
The perturbative expansion in tunneling in the many-lead case is valid when $\lambda_{ij}(E_C^*) \ll 1/M$, or $Mg\ll1$.

Upon reducing the cutoff to  $D\ll\Delta_{P}$, the dissipative action changes its coefficient, as discussed in the beginning of Sec.~\ref{subsec:2Weak-tunneling-limit}. 
Due to the changing coefficient of the dissipative part, the tunneling operator in the valley becomes marginally relevant; the action above changes to  
a one  identical to Eq.~(\ref{eq:SStrongSCvalley}), albeit with different UV cutoff now given by $\Delta_P$. 
The analysis is thus identical to the strong superconductivity case
discussed in Sec.~\ref{subsec:StrongSCCoulomb-valley} and we have
a topological Kondo effect. 
We can define a Kondo temperature that is valid in both regimes as 
\begin{equation}
T_{K}=\min(E_{C}^{*},\,\Delta_{P})\exp\left[-\frac{1}{(M-2)(g\Delta_{P}/E_{C}^{*})}\right]\,. \label{eq:TKGeneral}
\end{equation}
This estimate of the Kondo temperature  in the weak pairing regime $\Delta_{P}\ll E_{C}$ is one of our main results. 
It shows that the Kondo temperature is thus suppressed by the same exponential factor as in the case of strong pairing, $\Delta_{P}\gg E_{C}$. 
Equation~(\ref{eq:TKGeneral}) also gives the estimate $E_C^* \sim \min(T_K, M g \Delta_P)$ for the width of the charge degeneracy region. 

In the strong coupling limit, below the Kondo temperature, we find an irrelevant
reflection operator  $\sum_{j}r_{j}\cos(\mathbf{R}_{j}\cdot\boldsymbol{q}+2\frac{1}{M}\pi N_{g})$  with scaling dimension $\Delta=2(1-\frac{1}{M})$ similarly to Sec.~\ref{subsec:1Strong-tunneling-limit}, as expected from universality of the fixed point. 
(Strong bare tunneling in the weak pairing limit   will be further discussed in Sec.~\ref{subsec:2Strong-tunneling-limit}.) 
Despite weak bare tunneling and large charging energy, the low-temperature fixed point is characterized by strong tunneling (note similarity to  Ref.~\onlinecite{PhysRevB.96.041123}).
The pairing term which is a bulk term, unlike the boundary term $E_{C}$,
drives the island to Andreev-like fixed point of strong tunneling.

\subsubsection{Charge degeneracy point \label{subsec:WeakSCWeakTunChargeDeg}}

Near the charge degeneracy, $E_{C}^{*}\,,\Delta_{P} \ll D \ll E_{C}$, the
field $P$ is massless and the tunneling operator in Eq.~(\ref{eq:SdualWeakSC})
retains its scaling dimension $\Delta=1$. 
As before (Sec.~\ref{subsec:StrongScCD}), we  project on to two charge states of the island: 
$N^{+}\to S^{+}$, where the operator $S^{+} =S_x + i S_y =|1\rangle\langle0|$
acts on the  retained  charge states $0$ and $1$. 
The projected fermionic  action is then [compare to Eqs.~(\ref{eq:StFermion}) and~(\ref{eq:StrongSCHCK})]
\begin{equation}
S_{CK} \!= \!\! \int \! \! d\tau D \! \left[  \sum_{i}(\frac{J_{\perp}}{2} [s_i^- S^+ + s_i^+ S^-] \!+\!  J_z s_{z,i} S_z ) \! + \!  \gamma S_z \right]\! ,\label{eq:StFermionN}
\end{equation} 
where we introduced the pseudospin operators $s_i^- = (s_i^+)^\dagger = \bar{\psi}^{\dagger}_{i} \psi_{i} $ and $s_{z,i} = (\bar{\psi}^{\dagger}_{i} \bar{\psi}_{i}  -  \psi_{i}^{\dagger} \psi_{i})/2 $ and where $\bar{\psi}_{i}$ are the (neutral)  electron annihilation operators 
on the island at the contact $i$. 
The second  term, $J_z$, has a vanishing bare amplitude but will be generated upon reducing the bandwidth, see Eq.~(\ref{eq:WeakSCPeakRGJz}) below. 
The last term in Eq.~(\ref{eq:StFermionN}) describes finite detuning away from charge degeneracy with $\gamma=E_{C}^{*}/D$. 

Similar to Sec.~\ref{subsec:1Weak-tunneling-limit}, one obtains the weak-coupling charge Kondo RG equations~\cite{matveev1991quantum,1998PhRvB..57.5579Y,2002PhRvB..65s5101Y} ($J_{\perp}=2t_{i}$)
 \begin{flalign}
\frac{dJ_{\perp}}{dl} & =J_{\perp}J_{z}(1-\frac{M}{4}J_{z}) - \frac{M}{4} J_{\perp}^3\,,\label{eq:WeakSCPeakRG}\\
\frac{dJ_{z}}{dl} & =J_{\perp}^{2}(1-\frac{M}{2}J_{z}) \,. \label{eq:WeakSCPeakRGJz} \\
 \frac{d\gamma}{dl} & = \gamma (1- \frac{M}{4} J_{\perp}^2)\,. \label{eq:WeakSCPeakRGgamma}
\end{flalign}
The tunneling ($J_{\perp}$) operator is marginally relevant in the presence of
an infinitesimal $J_{z}$ term {[}compare to the ``superconducting''
case~\ref{subsec:1Weak-tunneling-limit}, Eq.~(\ref{eq:StrongSCPeakRG})
where the operator is relevant and $J_{z}$ is generated automatically{]}.
In the absence of superconductivity, there is an intermediate fixed
point~\cite{nozieres1980kondo} where $J_{\perp} \sim 1/M$.  
Equations~(\ref{eq:WeakSCPeakRG})--(\ref{eq:WeakSCPeakRGJz}) can be integrated by first identifying the constant of motion $(J_\perp^2 - J_z^2)/(1-\frac{M}{2}J_z)=$const. 
It is convenient to introduce $J_S = 1-\frac{M}{2}J_z$ and using the above relation obtain a closed equation for $J_S$~\cite{matveev1991quantum}. The solution can be obtained by solving the following equation:  
\begin{widetext}
\begin{align}
-\left. \frac{1}{4} M \left(\ln \frac{x^2}{xgM^{2}+4(x-1)^{2}} +\frac{ \left(g M^2-8\right) \tanh^{-1}[ \frac{\frac{1}{4}g M^2+2(x-1)}{\sqrt{g M^2} \sqrt{\frac{g M^2}{16}-1}}]}{2\sqrt{g M^2} \sqrt{\frac{g M^2}{16}-1}}\right) \right|_{x=J_S(0)}^{x=J_S(l)} = l \, ,    \label{eq:NormalCKSol}
\end{align}
\end{widetext}
where the initial value $J_S(0)=0$. 
Equation~(\ref{eq:NormalCKSol}) may be in principle inverted to find $J_S(l)$ and  $J_\perp(l) = \sqrt{J_S(l) g + \frac{4}{M^2}(1-J_S(l))^2}$. 

Let us first assume that the initial value of $J_\perp(0)\sim \sqrt{g}$ is below the intermediate fixed point value, $g \ll 1/M^2$. 
We then find from Eq.~(\ref{eq:NormalCKSol}), 
\begin{align}\label{eq:solJSNormal}
J_{S}(l)\approx \left \{ \begin{array}{cc}
                1-\frac{gM l}{2}, & |1-J_{S}(l)|  \ll 1 , \\
                \sqrt{\frac{4}{gM^{2}}}e^{-(\frac{2}{M}l-\frac{\pi}{\sqrt{gM^{2}}})} , & J_{S}(l) \ll 1 ,
               \end{array}
 \right.
\end{align}
The charge-Kondo temperature $T_{CK}$ can be obtained by solving the following equation $J_{\perp}(T_{CK})\sim 1/M$. One finds  $T_{CK} \sim E_{C} (gM^{2})^{\frac{M}{4}} e^{-\pi/(2\sqrt{g})}$~\cite{matveev1991quantum}. For simplicity we focus henceforth on $\Delta_P \gg T_{CK}$ limit. In this case, the charge-Kondo RG flow will be cutoff by $\Delta_P$ and at $D < \Delta_P$ Eqs.~(\ref{eq:WeakSCPeakRG})--(\ref{eq:WeakSCPeakRGJz}) should be replaced with Eqs.~(\ref{eq:StrongSCPeakRG})--(\ref{eq:StrongSCPeakRG2}).

Taking into account the solution \eqref{eq:solJSNormal}, one finds
$J^2_{\perp}(\Delta_P)\approx g(1-\frac{\pi}{4}\sqrt{gM^2}\frac{\ln (E_C/\Delta_P)}{\ln(E_C/T_{CK})}+\frac{\pi^2}{4}\frac{\ln^2 (E_C/\Delta_P)}{\ln^2(E_C/T_{CK})})$. The solution of topological Kondo RG equations can be obtained similarly as in Eq.\eqref{eq:solutionTK} to find that 
\begin{align}
T_K \sim M J_{\perp}(\Delta_P)^2 \Delta_P.   \label{eq:TkGeneral}
\end{align}
At this scale $J_{\perp}(T_K) \sim 1/ \sqrt{M}$. One can show that this scale (up to small logarithmic corrections) matches with the one obtained from the valley, Eq.~(\ref{eq:TKGeneral}) at $E_C^* \sim T_K$.

At lower energies $D\ll T_K$ the tunneling is strong so we must use perturbation theory in an irrelevant reflection operator, with a bare parameter $r(T_K) \sim 1$.  
In this regime  the findings of  Sec.~\ref{sec:StrongCoupling} apply since the low-temperature fixed point is universal:    
we have a leading irrelevant reflection operator   $\sum_{j}r_{j}\cos(\mathbf{R}_{j}\cdot\boldsymbol{q}+2\frac{1}{M}\pi N_{g})$ of scaling dimension $\Delta = 2(1- \frac{1}{M})$. 
The temperature-dependence of conductance has then the universal non-Fermi-liquid-type temperature-dependence, see Eq.~(\ref{eq:SummaryGBelowTk}).

Let us now consider $g \gg 1/M^2$ limit (still $g \ll 1/M$). 
We then find from Eq.~(\ref{eq:NormalCKSol}), 
\begin{align}\label{eq:solJSNormal}
J_{S}(l)\approx \left \{ \begin{array}{cc}
                1-\frac{gM}{2}l, &  |1-J_{S}(l)|  \ll 1 ,\\
           \frac{1}{gMl}  , & J_{S}(l)\ll 1 ,
               \end{array}
 \right.
\end{align}
The charge-Kondo temperature, obtained by solving $J_{\perp}(T_{CK}) \sim 1/ M$, is given by $T_{CK} \sim E_C e^{-2/(Mg)}$. 
As before we consider the limit $\Delta_P \gg T_{CK}$ here~\footnote{In the limit $\Delta_P \ll T_{CK}$ we reach the charge-Kondo fixed point and as a result have $J_{\perp}(\Delta_P) \sim 1/ M$.  In this case we find  $T_K \sim \Delta_P/M$.} and stop charge-Kondo RG flow at $\Delta_P$ finding $J^2_{\perp}(\Delta_P)\approx g(1-\frac{\ln (E_C/\Delta_P)}{\ln(E_C/T_{CK})})$. The corresponding Kondo temperature is given by Eq.~(\ref{eq:TkGeneral}).

The width of the charge degeneracy region can be estimated from the RG equation for the effective Zeeman coupling $\gamma$, Eq.~(\ref{eq:WeakSCPeakRGgamma}). 
When $J_{\perp} \ll 1/\sqrt{M}$, we can ignore the second term in the RHS of Eq.~(\ref{eq:WeakSCPeakRGgamma}) and use the approximate solution $\gamma = E_C^* / D$. 
This gives the width $E_C^*  \sim T_K$ for the charge degeneracy region, which agrees (up to logarithmic corrections) with our estimate from the valley, below Eq.~(\ref{eq:TKGeneral}). 

The perturbative weak-tunneling expansion is valid when $g \ll 1/M$, see  Sec.~\ref{subsec:WeakSCCoulomb-valley}. 
Remarkably, even in the cross over case $g \sim 1/M$, the above estimate [see below Eq.~(\ref{eq:TKGeneral})] gives a narrow width  $|\frac{1}{2}-N_{g}^{*}|\propto  \Delta_P / E_C \ll 1$ for the regime of strong charge fluctuations. 
This suggests that even if the bare conductance is large, Coulomb blockade oscillations of the conductance can form above the energy scale $\Delta_P$, see also Fig.~\ref{fig:gdiagram}. 
In the next section we confirm this expectation by studying the limit $g \gtrsim 1/M$ by using perturbation theory in reflection.

\subsection{Strong tunneling limit $Mg \gg 1$\label{subsec:2Strong-tunneling-limit}}

The strong tunneling limit in the weak pairing case is more complicated than in the strong pairing one discussed in Sec.~\ref{subsec:1Strong-tunneling-limit} because the reflection amplitudes have non-monotonic scaling behavior. Indeed, the reflection is relevant in the interval $\Delta_{P}\ll D\ll E_{C}$ but irrelevant for lower energies, $D\ll\Delta_{P}$. 
In particular, we will see that the conductance develops  Coulomb blockade oscillations at intermediate energies $\Delta_{P}\ll D \ll E_C$ when the  bare reflection amplitude $r = \sqrt{1-g}$ exceeds the cross over value $(\Delta_P / E_C)^{1/M}$. 
At the same time, the  charge of the island is not quantized but shows weak harmonic $N_g$-dependence. 

Let us  start from the strong-tunneling boundary action~(\ref{eq:SfullWeakSC}) 
in the limit $E_C \gg\Delta_{P}$. 
At $D\gg E_{C}$ we can  ignore the charging energy, finding that  the reflection operator is marginal,  $\Delta=\frac{1}{2}\frac{1}{2}|\mathbf{R}_{j}|^{2}+\frac{1}{2}2\frac{1}{M}=1$,  as expected for  free electrons. 
At energy scales below $E_{C}$, the field $Q$ becomes massive and can be integrated out.     
Averaging the action~(\ref{eq:SfullWeakSC}) over the massive field $Q$ and taking into account that $\langle e^{i2Q/\sqrt{M}}\rangle_Q \sim \left(\frac{E_{C}}{D}\right)^{1/M}e^{i2\pi N_{g}/M}$, we obtain in the range $\Delta_P < D< E_C$  [see Appendix~\ref{subsec:CorrFnQ}]
\begin{flalign}
 &S  =\frac{1}{\pi}T\sum_{i\omega_{n}}|\omega_{n}||\boldsymbol{q}(\omega_{n})|^{2}  -D\left(\frac{e^{\gamma}ME_{C}}{\pi D}\right)^{\frac{1}{M}} \nonumber \\
 &\times \int d\tau\sum_{j}r_{j}(E_{C})\cos \left(\mathbf{R}_{j}\cdot\boldsymbol{q}+2\pi\frac{1}{M}N_{g} \right) \label{eq:ActionWeakSCStrongTun0} \,.
\end{flalign}
As expected, the  scaling dimension of the resulting reflection operator has decreased,  
\begin{equation}
\Delta=\frac{1}{2}\frac{1}{2}|\mathbf{R}_{j}|^{2}=1-\frac{1}{M}<1\,. \label{eq:ScalingDimReflWSC}
\end{equation}
Thus backscattering is relevant  and we have the running amplitude  $r_{j}(D)=\left(\frac{E_{C}}{D}\right)^{\frac{1}{M}}  \left(\frac{e^{\gamma}M}{\pi}\right)^{\frac{1}{M}} r_{j}(E_{C})$. 
Since $r_j$ does not flow for $D\gg E_{C}$, we can fix  $r_{j}(E_{C})\sim\sqrt{1-g_{j}}$ where $g_{j}$ is the bare dimensionless conductance of the junction. 

Due to the finite weak reflection, the ground state energy has a weak harmonic  $N_g$-dependence. 
This dependence can be written in the form $\delta E_{GS}^{(M)}(N_g)=\Gamma \cos^2 \pi N_g$, where we introduce a renormalized charging energy~\cite{aleiner2002quantum}, 
\begin{equation}
   \Gamma\sim E_{C}r (E_{C})^{M}\,,\label{eq:Gamma}
\end{equation}
 see also  Eq.~\eqref{eq:WeakSCStrongTuncharge} and the discussion of the island average charge. Note that in the case $M=2$ one can find exact solution of the problem for $\Delta_P=0$~\cite{1995PhRvB..5216676F}. In the case of a finite p-wave gap, results depend on the ratio of $\Delta_P$ and $\Gamma$.

In the limit $\Gamma \ll\Delta_{P}\ll E_C$, the RG flow of $r_j(D)$ gets cutoff by $\Delta_P$ and reflection amplitude does not reach strong coupling. 
In the interval $E_{C}\gg D\gg\Delta_{P}$, we have weak reflection and weak pairing and we recover the results of Furusaki \& Matveev~\cite{PhysRevB.51.1743,matveev1994charge,1995PhRvB..5216676F}
(in the 2-lead case) and Yi \& Kane~\cite{1998PhRvB..57.5579Y,2002PhRvB..65s5101Y} (in the multi-lead case). 
At lower energies $\Gamma \ll D \ll \Delta_{P} $, fluctuations of  $\boldsymbol{q}$ in the island
become gapped. Thus, we find the following boundary action: 
\begin{flalign}
 &S \!= \! \frac{1}{2\pi}T\sum_{i\omega_{n}}|\omega_{n}||\boldsymbol{q}(\omega_{n})|^{2} -\! D \! \left(\frac{ D}{ \Delta_{P}}\right)^{1-\frac{2}{M}} \left(\frac{e^{\gamma}ME_{C} }{\pi \Delta_{P}}\right)^{\frac{1}{M}}\nonumber\\
 & \times \int d\tau\sum_{j}r_{j}(E_{C})\cos\left(\mathbf{R}_{j}\cdot\boldsymbol{q}+2\pi\frac{1}{M}N_{g}\right) \,,\label{eq:ActionWeakSCStrongTun}
\end{flalign}
One may notice that the reflection amplitude has a scaling dimension $\Delta=2(1-\frac{1}{M})>1$ and is irrelevant. Thus, the action~\eqref{eq:ActionWeakSCStrongTun} exhibits the universal strong coupling behavior of the topological Kondo effect with characteristic non-Fermi-liquid-type temperature-dependence of the conductance,  Eq.~(\ref{eq:SummaryGBelowTk}).

One the other hand, when $\Delta_{P}\ll\Gamma \ll  E_{C}$, reflection amplitudes grow under RG procedure until $r_j(\Gamma) \sim 1$. Thus, even though initially backscattering was weak the system eventually flows to the weak-tunneling regime. Therefore, we expect the case $\Delta_{P}\ll\Gamma < E_{C}$ to be analogous to the weak-tunneling regime,  
see Sec.~\ref{subsec:2Weak-tunneling-limit}. 
Given that $r_{j}(E_{C})\sim\sqrt{1-g_{j}}$, the scale $\Gamma$
becomes of order $E_{C}$ when $g\sim 1/M$ (up to logarithmic corrections in $M$). Thus,  the following discussion is limited to $Mg \gg 1$ case. 

At $\Delta_P \ll D\ll\Gamma$ we perform duality transformation and derive a tunneling action as explained in   Sec.~\ref{subsec:2Weak-tunneling-limit}. 
The dual action describes weak tunneling in the dual lattice of Eq.~(\ref{eq:ActionWeakSCStrongTun0}), i.e. hypertriangular lattice in the valley ($N_{g}=0$) and hyperhoneycomb
lattice at charge degeneracy ($N_{g}=1/2$). 

Let us consider first $N_g=0$ case. The dual action describing coherent electron transmission through the island is given in Eq.~\eqref{eq:SStrongSCvalley} with $\lambda_{ij}(D)=\lambda_{ij}(\Gamma)(D/\Gamma)$ for $\Delta_P \ll D\ll\Gamma$. Given that $r_{j}(\Gamma)\sim 1- O(\frac{1}{M})$ one can show, by matching the conductances at scale $\Gamma$ (see Eq.~(\ref{eq:GTunGamma})), that the corresponding cotunneling amplitude is $\lambda_{ij}(\Gamma) \sim 1/M$. 

When $D < \Delta_{P}$ modes in the island are gapped out and cotunneling becomes marginally relevant. Thus, we find the topological Kondo RG flow of $\lambda_{ij}$ with Kondo temperature

\begin{equation}
T_{K}=\Delta_{P} \exp\left[-\frac{1}{(M-2)\lambda(\Delta_P)}\right] , \label{eq:TKWeakScTun}
\end{equation}
with $\lambda(\Delta_P) \sim \Delta_{P}/(M \Gamma)$. \\

Let us now discuss the charge degeneracy case $N_g=1/2$. 
As before, at $\Delta_P \ll D\ll\Gamma$ we perform a duality transformation. 
The tunneling action is in this case~\cite{2002PhRvB..65s5101Y} 
\begin{equation}
    S_t \!=\!  
    -D\int d\tau\sum_{i} \frac{1}{2}  t_{i}(D) e^{i \frac{1}{2} \boldsymbol{R}_{i}\cdot\boldsymbol{p} } S^+ + H.c., \label{eq:Tunneling_Gamma}
\end{equation}
 where the operator $S^{+} =|1\rangle\langle0|$ acts on the  charge states $0$ and $1$.  $S^+$ can be also thought of as acting on the sublattice degree of freedom in the hyperhoneycomb lattice~\cite{2002PhRvB..65s5101Y}. 
 The scaling dimension of the tunneling operator is $\Delta = (1-\frac{1}{M})$ and together with its dual reflection operator, Eq.~(\ref{eq:ActionWeakSCStrongTun0}), the relation~\eqref{eq:Yi} is satisfied. 
 Thus, we obtain $t(D) \sim t(\Gamma) (\Gamma/D)^{1/M}$. 
 We can estimate the bare amplitude as $t(\Gamma) \sim 1/\sqrt{M}$ based on matching the conductances at scale $\Gamma$~\cite{2002PhRvB..65s5101Y}. 
 At $D < \Delta_{P}$ the modes in the island are gapped out and the scaling dimension of $t$  becomes $\Delta = (1-\frac{1}{M})$. Based on this, we find the Kondo temperature at charge degeneracy, 
 \begin{equation}
     T_K 
      \sim 
      \left(\frac{\Gamma}{\Delta_P}\right)^{\frac{2}{M+1}} \Delta_P \,, 
 \end{equation}
 obtained by solving from $t(T_K) \sim 1/ \sqrt{M}$.

Finally, at the lowest energies $D < T_K$ 
at any $N_g$, 
we  find an irrelevant reflection operator with scaling dimension $\Delta=2(1-\frac{1}{M})$.  
This corresponds to the  universal $T\to 0$ fixed point of the topological Kondo model. As before, the fractional scaling dimension leads to the non-Fermi-liquid-type temperature-dependence of the conductance, see Eq.~(\ref{eq:SummaryGBelowTk}).

\section{Charge and conductance\label{subsec:StrongSCAverage-charge-and}}

In this section we discuss in more detail the signatures of topological ground state degeneracy that were summarized in the introduction,  Sec.~\ref{sec:qualit}. 
We start from the thermodynamic
observable of average charge $\left\langle N\right\rangle$ and then discuss transport  signatures. 
We focus on temperatures $T \ll E_C$ so that we can ignore thermal fluctuations of the charge and concentrate on the relevant quantum effects. 

\begin{figure}[tb]
\includegraphics[width=1\columnwidth]{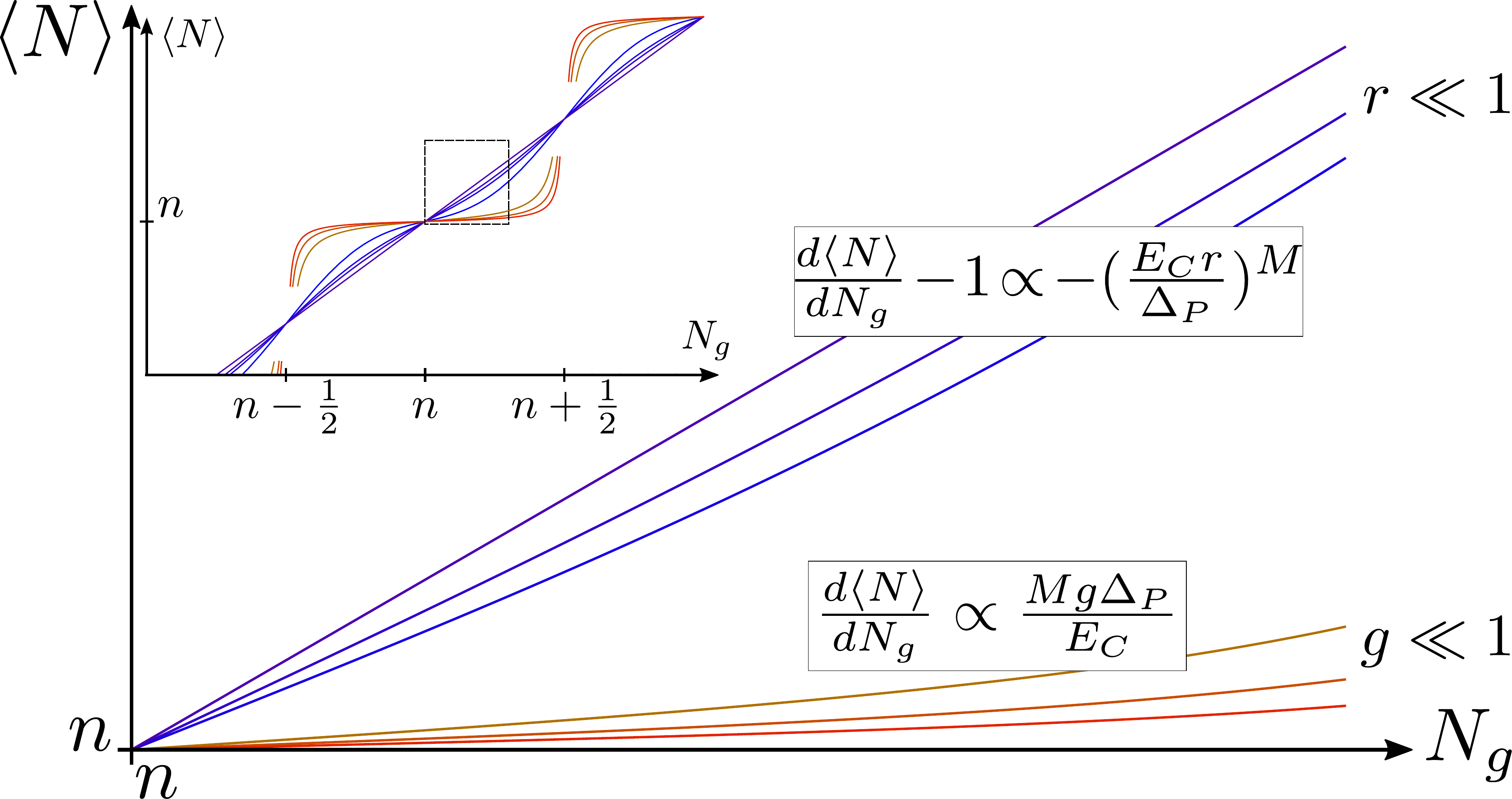}\caption{The island charge $\left\langle N\right\rangle$ vs gate charge $N_g$ in the strong superconductor ($\Delta_P \gg E_C$) regime. The red (nearly flat) curves correspond to weak tunneling ($M g \Delta_P\ll E_C$) while the blue ones correspond to weak reflection. 
\label{fig:charge}}
\end{figure}

\subsection{Average charge  in the strong superconductor limit $\Delta_P \gg E_C$ \label{sec:StrongSC_charge}}

In the weak tunneling limit, $M g \Delta_P \ll E_C$, the quantum fluctuations of charge are weak. 
The charge is then approximately quantized to the nearest   integer to the gate charge $N_g$,  $\left\langle N\right\rangle \approx \floor*{N_g + \frac{1}{2}}$, where $\floor*{\dots }$  denotes the floor function, see Appendix~\ref{sec:Average-chargeWT}. 
The charge steps at half-integer values of the gate charge are smeared out by thermal or quantum fluctuations~\cite{glazman1990lifting,matveev1991quantum}. 
At relatively high temperature, $T \gg M g \Delta_P $, the  broadening is thermal. 
Here we focus on  lower temperature at which  the broadening is of order $M g \Delta_P/E_C$  due to superconducting quantum fluctuations.  
The center of the charge plateau, $N_g =\text{integer}$, is horizontal in the absence of tunneling. In the presence of tunneling, it acquires a finite slope,  
$d\left\langle N\right\rangle/dN_g  \approx   \sum_{i} g_i  \Delta_P /E_C$, see Fig.~\ref{fig:charge} and Eq.~(\ref{eq:N2}) in Appendix~\ref{sec:Average-chargeWT}. 
The slope is enhanced by a factor $\Delta_P/E_C$ from the superconducting quantum fluctuations.  

In the strong tunneling limit, $M g \Delta_P \gtrsim E_C$,  
 the quantum fluctuations of charge are strong. 
 In this limit we can use perturbation theory in reflection~\cite{2016PhRvB..94l5407L}, see Sec.~\ref{subsec:1Strong-tunneling-limit}.  
In the absence of reflection, the ground state energy is independent
of gate charge since the island can compensate for $N_{g}$ by changing
its own charge; in the Hamiltonian one can shift $Q\to Q+\frac{1}{\sqrt{M}}\pi N_{g}$
which removes the $N_{g}$-dependence from it. 
Thus,  the average charge on the island equals the gate charge: $\left\langle N \right\rangle =N_{g}$. 
In the presence of reflection, the difference $\delta N \equiv \left\langle N \right\rangle - N_{g}$ becomes  non-zero (except for integer values of $N_g$). 
The lowest order contribution to  $\delta N$ starts at   order $M$ from a process where an electron is reflected from each contact, see  Appendix~\ref{sec:Average-chargeST}. 
When $M$ is large and $E_C /  \Delta_P \gg g \gg E_C / (M \Delta_P)$ we have $r(E_C)^M \sim (1 - \frac{g \Delta_P}{E_C})^M \ll 1$ 
and 
$\left\langle N\right\rangle -N_{g}\sim - (1 - \frac{g \Delta_P}{E_C})^M \sin2\pi N_{g}  $. 
In the limit $1 \gg g \gg E_C /  \Delta_P$, we find $r(E_C) \sim E_C / (g \Delta_P)$ and 
\begin{equation}
 \left\langle N\right\rangle -N_{g}\sim - \left(\frac{E_C}{ \Delta_P}\right)^{M} \left(\prod_{i=1}^M \frac{1}{g_i} \right) \sin2\pi N_{g} \,,  \label{eq:StrongSCStrongTuncharge1}
\end{equation} 
whereas when $g \approx 1 $, we have 
\begin{equation}
 \left\langle N\right\rangle -N_{g}\sim - \left(\frac{E_C}{\Delta_P}\right)^{M} \left(\prod_{i=1}^M (1-g_i)^{1/2}\right) \sin2\pi N_{g} \,.  \label{eq:StrongSCStrongTuncharge2}
\end{equation} 
Note that the amplitude of the harmonic corrections to $\left\langle N\right\rangle -N_{g}$  is weakened by a small factor $(E_C/\Delta_P)^M$ due to the enhanced superconducting charge fluctuations in topological superconductors. 
We also note that in Eq.~\eqref{eq:StrongSCStrongTuncharge2}  even a single fully transparent barrier, $g_i \to 1$, makes the  $N_{g}$-dependence fully linear~\cite{PhysRevLett.82.1245,2002PhRvB..66e4502F}. 
From Eqs.~(\ref{eq:StrongSCStrongTuncharge1})--(\ref{eq:StrongSCStrongTuncharge2}), the slope $d\left\langle N\right\rangle/dN_g $ near $N_g =\text{integer}$ is very close to 1,  $d \left\langle N\right\rangle /d N_g-1 \sim - (E_C/\Delta_P)^{M} r_1 \dots r_M $, see Fig.~\ref{fig:charge}. 
The  weak tunneling result for the slope matches with this estimate when $M g \Delta_P \sim E_C$.

\subsection{Average charge in the weak superconductor limit $\Delta_P \ll E_C$ \label{sec:weakSC_charge}}

In the case $E_C \gg \Delta_P$, weak superconductivity does not influence the average charge. 
The charge $\langle N\rangle$ behaves similarly as in normal metal islands since superconducting fluctuations are weaker than in the limit $E_C \ll \Delta_P$ discussed in the previous section. 
In the weak tunneling limit $g \ll 1/M$, we find Coulomb staircase with charge steps that are broadened by the charge-Kondo scale~\cite{matveev1991quantum} $T_{CK}/E_C$, see  Sec.~\ref{subsec:WeakSCWeakTunChargeDeg}. The slope  $d\left\langle N\right\rangle/dN_g $ is of order  $Mg$ due to weak tunneling, see Appendix~\ref{sec:Average-chargeWT}. 
In the strong tunneling regime $g \gtrsim 1/M$, we find~\cite{PhysRevLett.82.1245} (see  Appendix~\ref{sec:Average-chargeST})
\begin{equation}
 \left\langle N\right\rangle -N_{g}\sim - \left(\prod_{i=1}^M (1-g_i)^{1/2}\right) \sin2\pi N_{g} \,.  \label{eq:WeakSCStrongTuncharge}
\end{equation}
This result shows that the scale $\Gamma$ in Eq.~(\ref{eq:Gamma}) can be identified as a renormalized charging energy~\cite{aleiner2002quantum,2017PhRvL.119e7002L}. 

\subsection{Conductance in the strong superconductor limit $\Delta_P \gg E_C$ \label{sec:ConductanceStrongSC}}

\begin{figure}[tb]
\includegraphics[width=0.8\columnwidth]{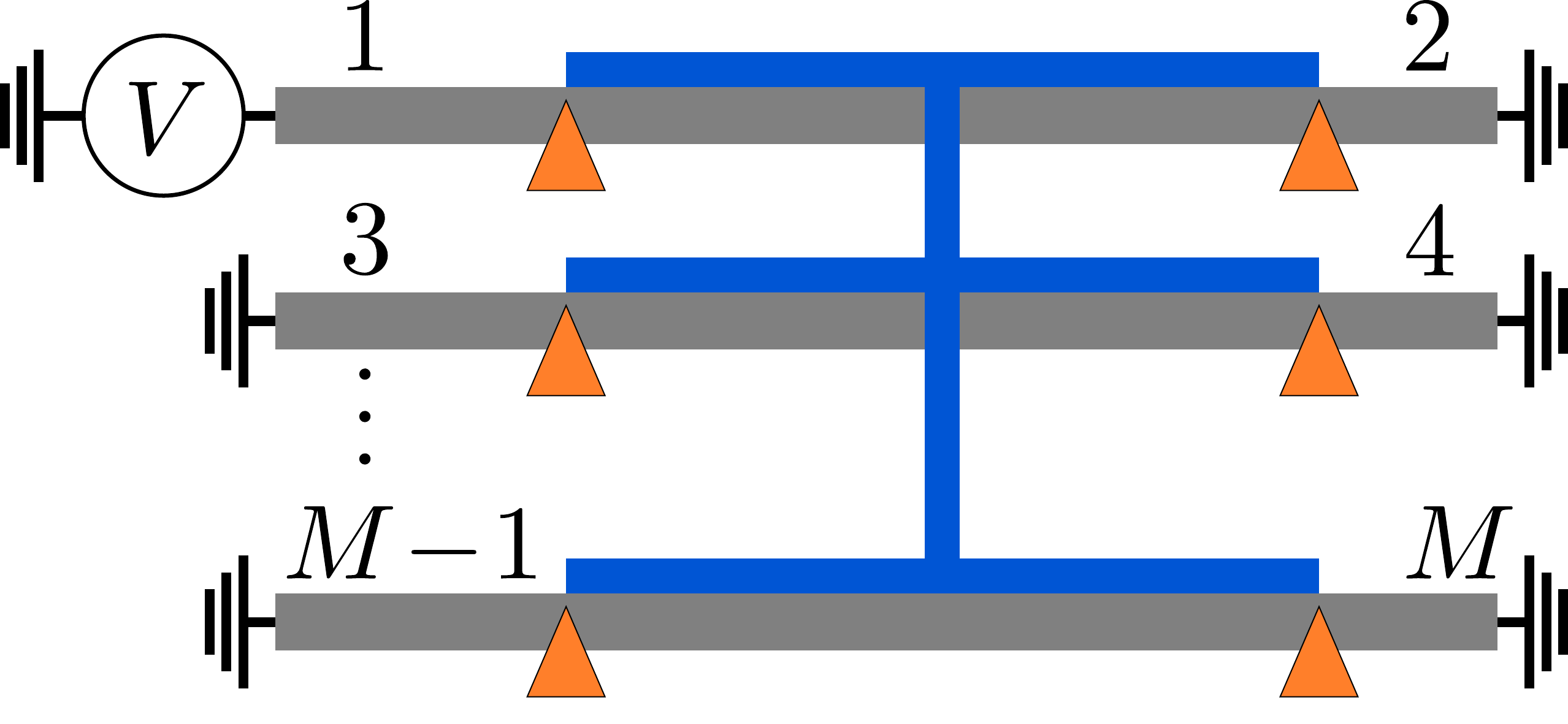}\caption{Schematic diagram of the biasing to measure the elements of the conductance matrix $G_{ij}$. \label{fig:device_bias}}
\end{figure}

The multiterminal island, Fig.~\ref{fig:device_bias}, is characterized by a conductance matrix $G_{ij}$, defined as   $\langle I_i \rangle  = \sum_{j} G_{ij} V_j$ which gives  the current $\langle I_i \rangle $ in contact $i$ as a response to voltage $V_j$ (measured from ground) applied to lead $j$.
(The current operator is $I_{i}=-e\frac{1}{\pi}\partial_{t}\varphi_{i}$.) 
The easiest way to measure the conductance matrix is to use an asymmetric biasing, see Fig.~\ref{fig:device_bias}. For example, one grounds all the other leads while applying  $V_j \neq 0$ in lead $j$. In this case, 
one measures the current in contact $i\neq j$
to get $G_{ij} = \langle I_{i} \rangle/ V_j $. The diagonal components of the conductance matrix are fixed by the conservation of current: $G_{ii} = -\sum_{j\neq i} G_{ij}$. 
The conductance matrix can then be calculated with the help of the Kubo formula, see Appendix~\ref{sec:currentop}.

\begin{figure}[tb]
\includegraphics[width=1\columnwidth]{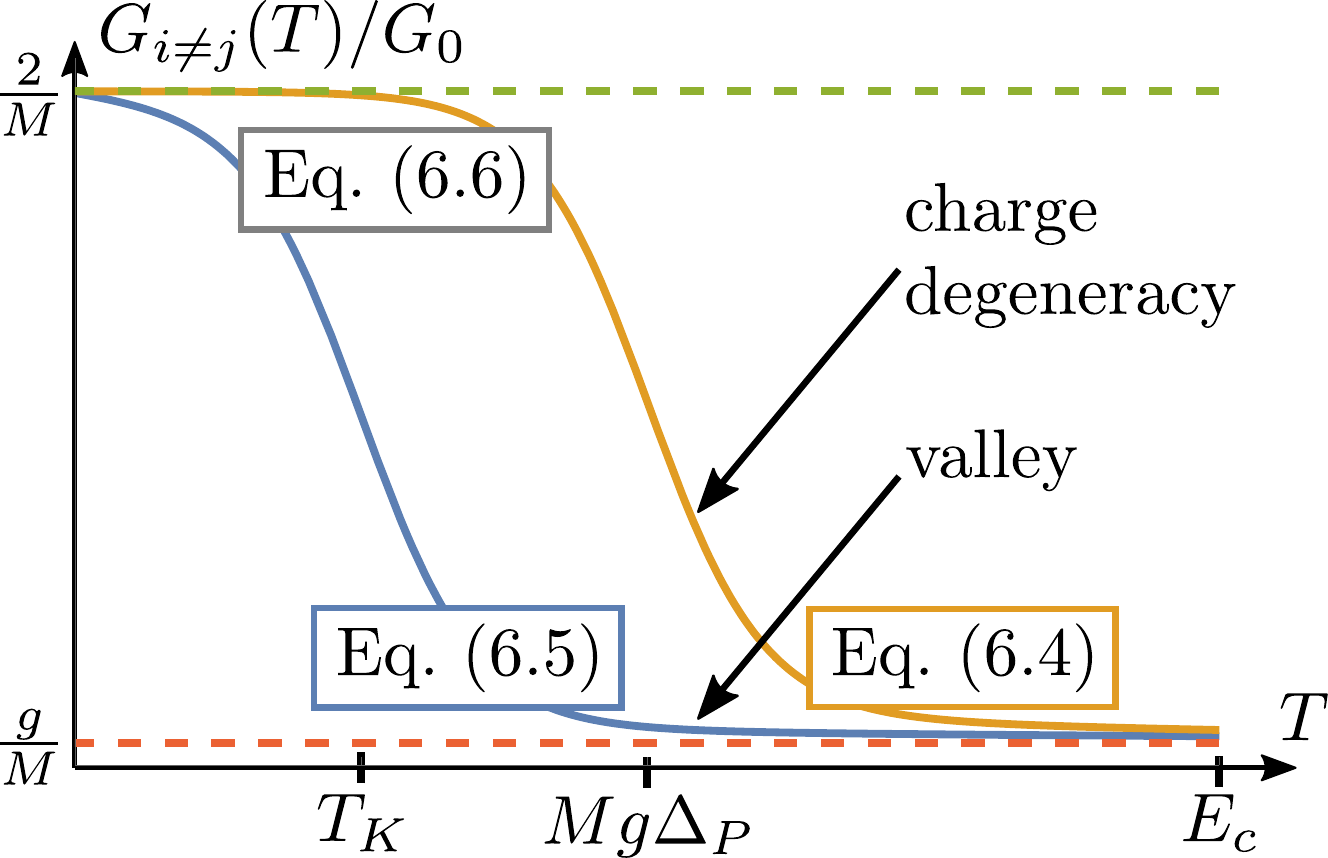}\caption{A sketch of the conductance vs $T$ in the strong superconductor and weak tunneling limit, $M g \Delta_P \ll E_C \ll \Delta_P$ in the Coulomb valley (blue, left) and at charge degeneracy (yellow, right). 
The scale $T_K$ of the topological Kondo effect is given by Eq.~(\ref{eq:TKStrongSC}). 
\label{fig:GStrongSCWeakTun}}
\end{figure}

Let us  start from the weak tunneling case, $M g \Delta_P \ll E_C$, Fig.~\ref{fig:regimes}a. 
The conductance in this regime as a function of temperature is sketched in  Fig.~\ref{fig:GStrongSCWeakTun}. 
At relatively high temperatures, $E_C \ll T \ll \Delta_P$, we have $G_{ij} \propto \delta_{ij}$. 
At   temperatures below the charging energy, $ T \lesssim E_C$, we have sequential tunneling through Majorana states,  yielding~\cite{2016PhRvB..93w5431V,2016arXiv160800581M}  
\begin{equation}
G_{ij}^{seq} =G_0  \frac{g_i (g_j - \delta_{ij} g_{\Sigma} )}{g_{\Sigma}} \frac{\Delta_P} { 4 T} \frac{1}{\cosh^{2} \frac{E_C^*}{2T}} \, \label{eq:Gseq} 
\end{equation}
with $g_{\Sigma} = \sum_i g_i$. 
The renormalization of junction conductances at $T \ll \Delta_P$ gave rise to the extra factor of $\Delta_P / T$ in Eq.~(\ref{eq:Gseq}). 
Away from charge degeneracy, sequential tunneling is suppressed by activation gap, $G^{seq} \propto e^{-E_C /T}$.

At temperature $T_K \ll T \ll E_C$, the conductance in the valley is instead dominated by elastic tunneling through a Majorana state, modified by the Kondo effect in the leading logarithmic approximation ($T>T_K$), 
\begin{equation}
\frac{G_{i  j} }{G_0} \approx \frac{\pi^{2}}{4} g_{i}\left(g_{j}-\delta_{ij}g_{\Sigma}\right) \left( \frac{\Delta_P} {E_C} \right)^2
\left[\frac{\ln (eE_C/T_K)  }{ \ln (eT / T_K)} \right]^2, \label{eq:GStrongSCKondo}
\end{equation}
 see Appendix~\ref{sec:currentopWeak}. Using the expression for Kondo temperature is $T_K \sim E_C e^{-E_C/Mg\Delta_P}$ at $N_g \approx 0$ one finds that $G_{i\neq j}/G_0 \sim  2 / M^2$ at  $T\to T_K$.

Below the Kondo temperature, we have a correction to the conductance from weak backscattering [see Eq.~(\ref{eq:GStrongCoupling})], 
\begin{equation}
\frac{G_{ij}}{G_0} \approx 2 \left(\frac{1}{M} - \delta_{ij}\right)  \left[ 1-  c\, r(T_K)^2 \left( \frac{T} {T_K} \right)^{2(1-\frac{2}{M})}  \right] \,, \label{eq:conductanceBelowTk}
\end{equation}
which displays the non-Fermi-liquid temperature-dependence. 
The numerical coefficient $c$ is beyond our RG analysis but expected to be of order unity. 
At $T \sim T_K$, we expect that $r(T_K) \sim 1 - O(1/M) $ and thus $G_{i\neq j}/G_0 \sim 1/M^2$. This matches with the limit $T \to T_K$ of the high-temperature result~(\ref{eq:GStrongSCKondo}).

Let us now discuss conductance through the island at $N_g \approx 1/2$, {\it i.e.} close to the charge degeneracy point. At high temperatures, the conductance is given by  Eq.~(\ref{eq:Gseq}) describing sequential tunneling through MZMs. At the charge degeneracy $E^*_C=0$ and, thus, $G_{i\neq j}/G_0 \approx   g \Delta_P / (4MT) $ for approximately isotropic barriers. At low temperatures $T\ll T_K$, the conductance is given by Eq.~\eqref{eq:conductanceBelowTk} with $T_K\sim Mg\Delta_P$. The two (high- and low-temperature) results match at $T\sim T_K$ yielding $G_{i\neq j}(T\sim T_K)/G_0 \sim 1/M^2$.  

In the strong-tunneling limit, $M g \Delta_P \gg E_C$ (Fig.~\ref{fig:regimes}b), the conductance has the same temperature-dependence as in  Eq.~(\ref{eq:conductanceBelowTk}) already at   relatively high temperatures $T \lesssim E_C$. Thus, the strong-tunneling regime may be favorable for observing the  non-Fermi-liquid temperature-dependence. 
We have for $T \ll E_C \ll\Delta_P$, (see Appendix~\ref{sec:currentopStrong}) 
\begin{equation}
G_{ij}/G_{0}=2\left(\frac{1}{M} - \delta_{ij}\right)
-c \left(\frac{T}{E_{C}}\right)^{2(1-\frac{2}{M})}
  R_{ij} \,, \label{eq:GStrongTun}
\end{equation}
where $c$ is a numerical coefficient whose calculation is beyond the accuracy of our RG treatment 
and the effective reflection coefficient is~\footnote{Apart from renormalization, the lowest order correction in $1-g_i$ to the sequential tunneling result $G^{seq}_{ij} \propto  \frac{g_i (g_j - \delta_{ij} g_{\Sigma} )}{g_{\Sigma}}$, Eq.~(\ref{eq:Gseq}), is also proportional to  $R_{ij}$.} 
\begin{equation}
R_{ij}=-\sum_{k}  r_k(E_C)^2 \left(\delta_{jk}-\frac{1}{M}\right)\left(\delta_{ik}-\frac{1}{M}\right) \,.     \label{eq:RijMT}
\end{equation}
For approximately  isotropic contacts, we have $R_{ij} \approx   \left(\frac{1}{M}- \delta_{ij}\right) r (E_C)^2$. 
From Sec.~\ref{subsec:1Strong-tunneling-limit}, we found that in the strong-tunneling limit  $r(E_C) \approx  E_C/(M g \Delta_P)$ when $1/M \gg g \gg E_C/ (M \Delta_P)$ and $r(E_C) =(E_C/{\Delta_P} ) \sqrt{1-g} $ when $g \gtrsim 1/M$. 
In the cross over between weak and strong tunneling we have $T_K \sim E_C$ and Eqs.~(\ref{eq:conductanceBelowTk}) and~(\ref{eq:GStrongTun}) match in that limit.

\subsection{Conductance in the weak superconductor limit $\Delta_P \ll E_C$ \label{sec:ConductanceWeakSC}}

In the weak superconductor limit, the dependence of conductance on temperature is non-monotonic. 
It is easy to see this effect in the strong tunneling regime, $g \gtrsim 1/M$ and $\Gamma \ll \Delta_P$ (Fig.~\ref{fig:regimes}d). Indeed, one finds in this case that the maximum of conductance doubles~\cite{2017PhRvL.119e7002L} from $\frac{e^2}{h} \frac{1}{M}$ to $\frac{2e^2}{h} \frac{1}{M}$ upon lowering the temperature across $T \sim \Delta_P$. 
Indeed, in the temperature interval $\Gamma , \Delta_P \ll T \ll E_C $, one finds~\cite{2002PhRvB..65s5101Y} 
the conductance from inelastic cotunneling of above-gap
quasiparticles, 
\begin{equation}
  G_{i j}/G_0 \approx \left(\frac{1}{M} - \delta_{ij}\right)  -  b_M 
    \left(\frac{E_{C}}{T}\right)^{\frac{2}{M}}
R_{ij} 
 \,, \label{eq:GWeakSCStrongTunAboveGap}
\end{equation}
where the coefficient $b_M$ is given in  Eq.~(\ref{eq:GStrongTunNormal}). 
The  increasing backscattering upon lowering the temperature agrees with  the RG flow of reflection amplitudes to scale $T$, see below Eq.~(\ref{eq:ScalingDimReflWSC}). 
The bare reflection amplitudes enter the coefficient $R_{ij}$, given in  Eq.~(\ref{eq:RijMT}) where now  $r_k(E_C)^2 \sim (1-g_k)$. 

The onset of p-wave  superconductivity at $T \sim \Delta_P$ modifies both the maximum values as well as temperature dependence. Indeed, for $T \ll \Delta_P$, the maximum conductance  is doubled~\cite{2017PhRvL.119e7002L},  
\begin{equation}
\frac{G_{i j}}{G_0} \! \approx 2 \left(\frac{1}{M} - \delta_{ij}\right) - c 
\left(\frac{E_C}{\Delta_P}\right)^{\frac{2}{M}} \left(\frac{T}{\Delta_P}\right)^{2(1- \frac{2}{M})} R_{ij} \,, \label{eq:GStrongTun2}    
\end{equation}
as the current is carried by the Majorana states, compare to  Eq.~(\ref{eq:GStrongTun}). 
Furthermore, unlike in Eq.~\eqref{eq:GWeakSCStrongTunAboveGap}, the backscattering corrections to conductance become \textit{smaller} as temperature is lowered, $dG_{i j}/dT < 0$. (Here $c$ is a constant of order one.) 
The conductance is therefore a non-monotonic function of temperature.~\footnote{The crossover form of the conductance doubling  is beyond this work but can be in principle estimated from the correction to Eq.~(\ref{eq:GWeakSCStrongTunAboveGap}) from perturbative p-wave pairing $\Delta_P$. }~\cite{PhysRevB.96.041123} 
The backscattering corrections to Eqs.~(\ref{eq:GWeakSCStrongTunAboveGap})--(\ref{eq:GStrongTun2}) match at $T \sim \Delta_P$. 


The non-monotonic dependence of the conductance vs temperature is also present in the weak tunneling limit,  $\Gamma \gg \Delta_P$ [corresponding to $g \ll 1-  (\Delta_P/E_C)^{2/M}]$. 
  
In the Coulomb valleys,  the conductance becomes small [see Eq.~(\ref{eq:AppGTunGamma})] at temperatures $\Delta_P \ll T  \ll \Gamma$,  
\begin{equation}
G_{ij}/G_0  =  \left(\lambda_{ij}(\Gamma)^2 -\delta_{ij} \sum_k  \lambda_{ik}(\Gamma)^2\right)\left(\frac{ T}{\Gamma}\right)^{2}\,, \label{eq:GTunGamma}
\end{equation}
 We can match Eq.~(\ref{eq:GTunGamma})  with the strong-tunneling result  Eq.~(\ref{eq:GWeakSCStrongTunAboveGap}) at $T\sim \Gamma$. In the latter, the backscattering correction becomes $b_M (E_C/\Gamma)^{2/M} R_{ij} \sim \frac{1}{M} +O(\frac{1}{M^2}) $ and thus $G_{i\neq j} \sim 1/M^2$ which matches with  Eq.~(\ref{eq:GTunGamma})  when $\lambda(\Gamma) \sim 1/M$. (Here we neglect $\ln M$ corrections.)
 

Upon lowering the temperature to $T\ll \Delta_P$, the tunneling in the valley becomes marginally relevant leading to a weak-coupling Kondo RG flow with characteristic scale $T_K \sim  \Delta_{P}e^{-\Gamma/\Delta_{P}}$, see Eq.~(\ref{eq:TKWeakScTun}). 
The conductance in this case reads
\begin{equation}
\frac{G_{i  j} }{G_0} \! \approx   \left[\lambda_{ij}(\Gamma)^2 \!-\delta_{ij}\sum_k \lambda_{ik}(\Gamma)^2 \right] \! \left[ \frac{\Delta_P} {\Gamma} \right]^2
\left[\frac{\ln \frac{e\Delta_P}{T_K}  }{ \ln \frac{e T}{T_K}} \right]^2 \! , \label{eq:GWeakSCKondo}
\end{equation}
where we assumed that $T_K \ll T \ll \Delta_P$ and $\lambda(\Gamma) \sim 1/M$. 
At $T \sim \Delta_P$, the expressions~(\ref{eq:GWeakSCKondo}) and~(\ref{eq:GTunGamma}) match. 
At $T\to T_K$, the logarithmic factor becomes of order  $ \sim (\Gamma/\Delta_P)^2$ and we find  $G_{i\neq j}/G_0 \sim  1 / M^2  $.  
Finally, in the low temperature regime  $T \ll T_K$, the non-local conductance approaches the quantized value $2G_0/M$, see ~(\ref{eq:conductanceBelowTk}).  
At $T \ll T_K $, the conductance is given by Eq.~(\ref{eq:conductanceBelowTk}) with the aforementioned $T_K$. One can see that this results matches with Eq.~(\ref{eq:GWeakSCKondo}) at $T\sim T_K$.

\section{On the relation to the multi-channel Kondo model \label{sec:compare}}

Since in our model we include the superconducting pairing explicitly, we can understand the relation between the topological Kondo effect and the earlier studies of 
inelastic cotunneling in multi-terminal normal state metallic quantum dots (which can be mapped to the multichannel Kondo problem) in Refs.~\onlinecite{matveev1991quantum,1995PhRvB..5216676F,1998PhRvB..57.5579Y,2002PhRvB..65s5101Y}. 

We demonstrated that both the topological superconducting as well as normal state models can be mapped to a model of Brownian motion of a quantum particle in a dissipative bath. 
The fictitious quantum particle describes the charge that tunnels across a tunnel barrier~\cite{1992PhRvL..68.1220K} between the lead and the island, while the bath is formed by low-energy electronic fluctuations away from the tunneling contact.  
The main difference between normal state and topological superconducting
island is in the effective dissipation strength (i.e. effective Luttinger liquid parameter) in the  quantum Brownian motion model, which in turn determines the scaling dimensions of boundary operators. 
The difference stems from the fact that in the case of a topological superconducting island, the fluctuations away from the tunnel contact are created only on half-axis since the superconducting region is gapped out at energies below  
$\Delta_P$. 
This modifies the dissipation strength (and all scaling dimensions) by a factor 2 because the superconducting region does not contribute to dissipation. 

In the normal state, the multi-terminal quantum dot model at charge degeneracy can be mapped to the multichannel Kondo (MCK) model. 
The multichannel Kondo model  has a stable intermediate fixed point~\cite{nozieres1980kondo} characterized by a universal conductance $G_{ij}^{MCK} = \frac{e^2}{h} \left( \frac{1}{M} - \delta_{ij} \right) \sin^2 \frac{\pi}{M+2}$ at half-integer values of $N_g$. (Away from these values, the conductance is small.)  
Due to the difference in dissipation strength, the topological Kondo model does not have an intermediate fixed point for strictly non-interacting leads. 
The  low-temperature fixed point 
is characterized by a conductance
$G_{ij} = \frac{e^2}{h} \left( \frac{1}{M} - \delta_{ij} \right) $ which is independent of $N_g$.  

\section{Conclusions \label{sec:Conclusions}}

We study signatures of topological ground-state degeneracy in a mesoscopic Majorana island. We have developed a microscopic model which explicitly includes p-wave superconductivity and have identified the signatures of ground-state degeneracy in thermodynamic and transport observable quantities. 

Our work sheds light on the so-called topological Kondo effect which corresponds to a formation of a correlated state between an effective Majorana ``spin" in the island and electrons in the leads. Our model allows one to express the characteristic Kondo scale $T_K$ in terms of microscopic parameters: junction conductances $g_i$, charging energy $E_C$ and topological gap $\Delta_P$.  We have also extended the previous results for the topological Kondo effect to the regimes of weak pairing ($\Delta_P \ll E_C$) as well as to strong tunneling limit ($g \sim 1$). Thus, our results provide insight regarding the experimental requirements necessary to detect this exotic correlated state.

We find that the main features of the topological Kondo effect, the quantized conductance and its non-Fermi-liquid-like  temperature-dependence, are present in both limits: $\Delta_P \gg E_C$ and $\Delta_P \ll E_C$. 
In the strong tunneling limit, the effective Kondo temperature  becomes large, $T_K \sim \min(E_C,\Delta_P)$,  
which makes the temperature-window for observing non-Fermi-liquid corrections to conductance in the laboratory favorable. 
The underlying reason for the robustness of the topological Kondo effect stems from the fact that the topological ground-state degeneracy is protected by the bulk gap (i.e. superconducting p-wave pairing is a relevant bulk perturbation), thus leading to the same universal low-temperature fixed point regardless of $E_C/\Delta_P$. The non-Fermi-liquid fixed point is stable and isotropic as long as the hybridization between different Majoranas can be ignored~\cite{2014JPhA...47z5001A}. 

In islands with more than two (non-interacting) leads, the superconducting fluctuations are enhanced due to ground-state degeneracy and, as a result, charging effects are suppressed. This effect is particularly dramatic when $\Delta_P \gg E_C$, in which case we find that charging energy has additional suppression relative to the normal-state Coulomb islands,  proportional to $(E_C/\Delta_P)^M$. This renormalization of the charging energy can be used to identify ground-state degeneracy. The suppression of charging energy is also important in the context of topological qubits where charging energy provides protection from quasiparticle poisoning~\cite{2017NJPh...19a2001P,2017PhRvB..95w5305K}.

\begin{acknowledgments}
We thank Leonid Glazman, Yuval Oreg,  Masaki Oshikawa, Dmitry Pikulin, and  Eran Sela for discussions.  This work was performed in part at Aspen Center for Physics, which is supported by National Science Foundation grant PHY-1607611. 
AEF was supported by the US Department of Energy (DOE), Office of Science, Basic Energy Sciences grant number DE-SC0019275. 
\end{acknowledgments}

\appendix

\begin{widetext}

	\section{Derivation of the dissipative action~(\ref{eq:Sfull0}) \label{sec:More-detailed-model}}
	In this Appendix we discuss how p-wave pairing term~(\ref{eq:HP}) modifies the dissipation strength in the effective boundary model. 
	For this, it is convenient to use the fermion representation. 
	We linearize the spectrum  around the Fermi level, which is valid at small energies compared to the Fermi energy. 
	The Hamiltonian in the proximitized segment is then 
		\begin{equation}
		H_{island}=\sum_{\alpha=1}^{M/2} \int_{0}^{L}dx \left(\sum_{r=R,L} \psi_{\alpha r}^{\dagger}(-ivr\partial_{x}-vk_{F})\psi_{\alpha r}+[\Delta_{P}e^{2i\theta_{\text{SC}}}\psi_{\alpha R}^{\dagger}\psi_{\alpha L}^{\dagger}+h.c.]\right)\,,
		\end{equation}
		where $\psi_{\alpha r}$ is a spinless fermion operator for wire $\alpha$
		and $r=+,-=R,L$ labels right and left movers.
		We will suppress the label $\alpha$ in the remainder of this section. 
		In the pairing term, the operator $e^{i\theta_{\text{SC}}}$ removes an electron from the superconductor backbone, $[N_{\text{SC}},e^{-i\theta_{\text{SC}}}]=1$.
		The total charge (that appears in the charging energy) $N=N_{\text{SC}}+\sum_{\alpha}N_{\alpha}$
		is thereby conserved by the pairing term. We can diagonalize $H_{island}$ 
		with a bogoliubov transformation in momentum space, 
		\begin{equation}
		c_{rk,r}=e^{i\theta_{\text{SC}}}(d_{kr}u_{k}-rv_{k}d_{k\overline{r}}^{\dagger})\,,
		\end{equation}
		where $d_{kr}$ is a neutral operator {[}commutes with $N${]} and
		\begin{equation}
		u_{k}=\frac{1}{\sqrt{2}}\sqrt{1+\frac{\varepsilon_{k}}{\sqrt{\Delta_{P}^{2}+\varepsilon_{k}^{2}}}}\,,\quad v_{k}=\frac{1}{\sqrt{2}}\sqrt{1-\frac{\varepsilon_{k}}{\sqrt{\Delta_{P}^{2}+\varepsilon_{k}^{2}}}}\,,
		\end{equation}
		with $\varepsilon_{k}=v(k-k_{F})$. The diagonalized Hamiltonian is
		$H_{island}=\sum_{\alpha, r}\sum_{k}\sqrt{\Delta_{P}^{2}+\varepsilon_{k}^{2}}d_{k,\alpha,r}^{\dagger}d_{k,\alpha,r}$. 
		Note that the phase $\theta_{\text{SC}}$ does not enter in the backscattering operator $\psi_{R}^{\dagger}\psi_{ L}$. 
		
		
		By using the diagonalized Hamiltonian we find that $\left\langle [\psi_{R}^{\dagger}\psi_{L}](\tau)[\psi_{L}^{\dagger}\psi_{R}](0)\right\rangle \sim \tau^{-2} e^{-2\tau\ensuremath{\Delta_{P}}}$
		in the proximitized segment. In the bosonic description, $\psi_{r}(x)=e^{i(r\varphi-\theta)}$, this corresponds
		to a non-dissipative effective action at low frequencies   $\frac{1}{\pi}T\sum_{\omega_{n}} e^{-|\omega_n|/D} \frac{|\omega_{n}|^{2}}{\Delta}|\varphi(x,\omega_{n})|^{2}$  (with $D\ll\Delta_P$)   
		for $0 \ll x \ll L$ 
		after integrating out the spatial fluctuations. Thus, the bosonic
		action at the boundary ($x=0,L$) of the proximitized segment is dominated by
		the dissipative part, $\frac{1}{2\pi}T\sum_{\omega_{n}}\sum_{j}|\omega_{n}||\varphi_{j}(\omega_{n})|^{2}$, the first term  in Eq.~(\ref{eq:Seff0}). 

\section{Derivation of the dual action by using Villain transformation\label{sec:Detailed-derivation-of}}
In this Appendix we derive the dual boundary action~(\ref{eq:Sdual}) used in the weak tunneling limit. 
We start from the boundary action~(\ref{eq:Sfull0}) in the limit of large barriers $Dr_{i}$. 
We consider the boundary-field 
partition function, 
\begin{flalign}
Z & =\int \mathcal{D}\varphi_{1}\dots \mathcal{D}\varphi_{M}\label{eq:Zphis}\\
 & \exp-\left(\frac{1}{\pi}T\sum_{\omega_{n}}\sum_{i}\frac{1}{K}|\omega_{n}||\varphi_{i}(\omega_{n})|^{2}+\int d\tau\sum_{i}D r_{i}\cos2\varphi_{i}(\tau)+\frac{1}{\pi^{2}}E_{C}\int d\tau(\sum_{i}\varphi_{i}(\tau)-\pi N_{g})^{2}\right)\,.
\end{flalign}
We include a factor $K$ which takes values 2 or 1 in the cases where the cutoff is either much below $\Delta_P$, used in the strong pairing case Eq.~(\ref{eq:Sfull0}), or much above it, used in the weak-pairing limit, Eq.~(\ref{eq:SfullWeakSC}). 

Because of the large $Dr_i$, we can approximate the cosine as~\cite{villain1975theory,PhysRevB.85.245121} $\cos x_{i}\approx\sum_{n_{i}}\frac{1}{2}(x_{i}-(2n_{i}+1)\pi)^{2}$
which leads to 
\begin{equation}
Z\approx\int \mathcal{D}\varphi_{1}\dots \mathcal{D}\varphi_{M}e^{-S_{0}}\sum_{n_{1}(\tau)\dots n_{M}(\tau)}e^{-S_{pot}}
\end{equation}
where $S_{0}=\frac{1}{\pi}T\sum_{i}\sum_{\omega_{n}} \frac{1}{K}|\omega_{n}||\varphi_{i}(\omega_{n})|^{2}$
and 
\begin{flalign}
S_{pot} & =2 \int d\tau\sum_{i}Dr_{i}(\varphi_{i}-a_{i})^{2}+\frac{1}{\pi^{2}}E_{C}\int d\tau(\sum_{i}\varphi_{i}-\pi N_{g})^{2}
\end{flalign}
and we denote $a_{i}=\frac{1}{2}(2n_{i}+1)\pi$. 

Next, we introduce the decoupling fields $z_j$, 
\begin{flalign}
e^{-S_{pot}} & =e^{-\frac{1}{\pi^{2}}E_{C}\int d\tau(\sum_{i}\varphi_{i}-\pi N_{g})^{2}}\int \mathcal{D}z_{1}\dots \mathcal{D}z_{M}\exp-\int d\tau\sum_{j}\left(\frac{1}{2Dr_{j}}z_{j}^{2}-2iz_{j}(\varphi_{j}-a_{j})\right)\\
 & =e^{-\frac{1}{\pi^{2}}E_{C}\int d\tau(\sum_{i}\varphi_{i}-\pi N_{g})^{2}}\int \mathcal{D}{\delta \theta}_{1}\dots \mathcal{D} {\delta \theta}_{M}\exp-\int d\tau\sum_{j}\left(\frac{1}{2Dr_{j}}\frac{1}{4\pi^{2}}(\partial_{\tau} {{\delta \theta}}_{j})^{2}-i\frac{1}{\pi}(\partial_{\tau} {\delta \theta}_{j})(\varphi_{j}-a_{j})\right)\,. 
\end{flalign}
We set in the second line $z_{j}=\frac{1}{2\pi}\partial_{\tau} {\delta \theta}_{j}$ where
${\delta \theta}_{j}$ are conjugate to $\frac{1}{\pi}\varphi_{i}$: $[\frac{1}{\pi}\varphi_{j},{\delta \theta}_{j}]=i$.
We will drop the terms $\propto 1/Dr_j$ from hereon. 

Only $\partial_{\tau} \delta \theta_{j}$ is now coupled to the integer-field $n_{j}$ so we can do the sum. 
This part of the path integral reads 
\begin{equation}
\sum_{n_{1}(\tau)\dots n_{M}(\tau)}\exp-\int d\tau\sum_{j}i\frac{1}{2\pi} \partial_{\tau} \delta \theta_{j}  n_{j}2\pi \,.
\end{equation}
Upon inspecting this with the Poisson summation formula, $\sum_{s\in\mathbb{Z}}\delta(s-{\delta \theta})=\sum_{m\in\mathbb{Z}}e^{2\pi im{\delta \theta}}$,
the sum over $n_{i}(\tau)$ imposes $\frac{1}{2\pi}{\delta \theta}_{i}$
to be integer-valued. We impose this condition ``softly'' by introducing a cosine potential~\cite{PhysRevB.85.245121}, 
\begin{equation}
\sum_{n_{1}(\tau)\dots n_{M}(\tau)}e^{-\int d\tau\sum_{j}i\frac{1}{2\pi} \partial_{\tau} \delta \theta_{j}n_{j}2\pi} \approx e^{\int d\tau\sum_{j}D t_{j}\cos{\delta \theta}_{j}} \,.
\end{equation}

Next, we rotate to the basis~(\ref{eq:phiqbasis})  where the total charge mode
$\sum_{i}\varphi_{i}$ is singled out. Similar to the basis change~(\ref{eq:phiqbasis}), we transform the conjugate fields as 
\begin{equation}
{\delta \theta}_{i}=\frac{1}{2}\boldsymbol{R}_{i}\cdot\boldsymbol{p}+\frac{1}{\sqrt{M}}P \,.
\end{equation}
Then 
$[q_j, p_k]=i\pi \delta_{jk}$ and  $[Q,P]=i\pi$. In the new variables, the full action is (we shift the $Q$ field
$Q\to Q+\frac{\pi}{\sqrt{M}}N_{g}$) 
\begin{flalign}
S[\boldsymbol{q},Q,\boldsymbol{p},P] & =\frac{T}{\pi}\sum_{\omega_{n}}\left[\frac{1}{K}|\omega_{n}||\boldsymbol{q}(\omega_{n})|^{2}+\frac{1}{K}\left(|\omega_{n}|+\frac{KM}{\pi}E_{C}\right)|Q(\omega_{n})|^{2}\right]\! -\int d\tau i \partial_\tau P \frac{1}{\sqrt{M}}N_{g} \\
 & 
 -\! \frac{1}{\pi}\int d\tau\sum_{j}\left(i\frac{1}{4}(\boldsymbol{R}_{j}\cdot \partial_\tau \boldsymbol{p})(\boldsymbol{R}_{j}\cdot\boldsymbol{q})+i\frac{1}{M} \partial_\tau P Q\right) 
 -\int d\tau\sum_{i}D t_{i}\cos\left(\frac{1}{2}\boldsymbol{R}_{i}\cdot\boldsymbol{p}+\frac{1}{\sqrt{M}}P\right)\,.
\end{flalign}
In general, $P(\tau+\frac{1}{T})=P(\tau)+2\pi\sqrt{M}W$ with $W$ an integer
winding number. Let us define $P(\tau)=\overline{P}(\tau)+2\pi\sqrt{M}WT\tau$
where $\overline{P}$ is $1/T$-periodic. Then, 
\begin{flalign}
S[\boldsymbol{q},Q,\boldsymbol{p},\overline{P}] & =\frac{T}{\pi}\sum_{\omega_{n}}\left[\frac{1}{K}|\omega_{n}||\boldsymbol{q}(\omega_{n})|^{2}+\frac{1}{K}\left(|\omega_{n}|+\frac{KM}{\pi}E_{C}\right)|Q(\omega_{n})|^{2}\right]\\
 &-\int d\tau\sum_{j}\frac{i}{\pi}\left(\frac{1}{4}(\boldsymbol{R}_{j}\cdot \partial_\tau \boldsymbol{p})(\boldsymbol{R}_{j}\cdot\boldsymbol{q})+\frac{1}{M} \partial_\tau \overline{P}Q+\frac{1}{M}2\pi\sqrt{M}WTQ\right) \\
 &-i2\pi WN_{g}  -\int d\tau\sum_{i}D t_{i}\cos\left(\frac{1}{2}\boldsymbol{R}_{i}\cdot\boldsymbol{p}+\frac{1}{\sqrt{M}}\overline{P}+2\pi WT\tau\right) \,.
\end{flalign}
We will from hereon drop the bar from $\overline{P}$. 
Integrating out $q$ and $Q$, we are left with the final dual partition function 
\begin{flalign}
Z & =\sum_{W}e^{iN_{g}2\pi W-\pi^{2}TE_{C}^{-1}W^{2}} \int \mathcal{D} \boldsymbol{p} \mathcal{D} P \exp-\frac{T}{4\pi }\sum_{\omega_{n}} K |\omega_{n}|\left\{ |\boldsymbol{p}(\omega_{n})|^{2}+\frac{|\omega_{n}|}{|\omega_{n}|+\frac{KM}{\pi}E_{C}}|P(\omega_{n})|^{2}\right\} \\
 & \exp\int d\tau\sum_{i}D t_{i}\cos\left(\frac{1}{2}\boldsymbol{R}_{i}\cdot\boldsymbol{p}+\frac{1}{\sqrt{M}}P+2\pi WT\tau\right) \,, \label{eq:ZDual}
\end{flalign}
which gives the actions~(\ref{eq:Sdual}) and~(\ref{eq:SdualWeakSC}) of the main text, by setting respectively $K=2$ or $K=1$ above.

\subsection{Derivation of the tunneling action in Sec.~\ref{subsec:1Weak-tunneling-limit} \label{sec:TunAction}}

Expanding Eq.~(\ref{eq:ZDual}) to second order in the tunneling amplitudes $t_i$, we find (we take $K = 2$, relevant for the strong superconductor limit of Sec.~\ref{sec:Strong-superconductor-limit}) 
\begin{flalign}
Z & \approx\int \mathcal{D} \boldsymbol{p} e^{-\frac{T}{2\pi}\sum_{i\omega_{n}}|\omega_{n}||\boldsymbol{p}(\omega_{n})|^{2}}\sum_{W}e^{iN_{g}2\pi W-\pi^{2}TE_{C}^{-1}W^{2}}
\int \mathcal{D} P e^{-\frac{T}{2\pi}\sum_{i\omega_{n}}
\frac{|\omega_{n}|^{2}}{|\omega_{n}|+\frac{2M}{\pi}E_{C}}
|P(\omega_{n})|^{2}}\left(1+\frac{1}{2}S_{t}^{(W)2}\right)\\
 & \approx Z_{0}(N_{g})Z_{P}\int \mathcal{D} \boldsymbol{p} e^{-\frac{T}{2\pi}\sum_{i\omega_{n}}|\omega_{n}||\boldsymbol{p}(\omega_{n})|^{2}}e^{-S_{t,\text{eff}}} \,,
\end{flalign}
where we denote $S_{t,\text{eff}}=-\frac{1}{2}\left\langle S_{t}^{(W)2}\right\rangle _{P,W}$
and 
\begin{equation}
Z_{0}(N_{g})  =\sum_{W=-\infty}^{\infty}e^{iN_{g}2\pi W-\pi^{2}TE_{C}^{-1}W^{2}}=\vartheta(N_{g};\,i\pi TE_{C}^{-1})\,, \quad 
Z_{P}  =\int \mathcal{D} Pe^{-\frac{T}{2\pi}\sum_{i\omega_{n}}\frac{|\omega_{n}|^{2}}{|\omega_{n}|+\frac{1}{\pi}2ME_{C}}|P(\omega_{n})|^{2}} \,, \label{eq:Zp}
\end{equation}
and we dropped the first order term $\left\langle S_{t}^{(W)}\right\rangle _{P,W}$
which is negligible. 

The partition function  $Z_{0}(N_{g})$ gives the charge steps of an isolated island, see Sec.~\ref{subsec:Ground-state-energy}; $\vartheta$ is the Jacobi theta function.
The effective tunneling action~(\ref{eq:StEffStrongSC}) used in Sec.~\ref{subsec:1Weak-tunneling-limit}  is obtained from 
\begin{equation}
S_{t,\text{eff}}  =-\frac{1}{4} D^{2}\int d\tau\int d\tau' \left\langle e^{i\frac{1}{\sqrt{M}}[P(\tau)-P(\tau')]+2\pi iWT(\tau-\tau')}\right\rangle_{P,W}\sum_{ij}t_{i}(D)t_{j}(D) 
\cos \left(\frac{1}{2}\boldsymbol{R}_{i}\cdot\boldsymbol{p}(\tau)-\frac{1}{2}\boldsymbol{R}_{j}\cdot\boldsymbol{p}(\tau')\right), \label{eq:Steff}
\end{equation}
by integrating out $P$ and summing over $W$. The correlation function in the integrand is discussed in Sec.~\ref{subsec:The-correlation-function}.

\section{Vertex operator averages and correlation functions \label{sec:CorrFn}}

In this Section, we calculate the averages  and   correlation functions of the boundary vertex operators $ e^{2i\frac{1}{\sqrt{M}}Q}$, $ e^{i\boldsymbol{R}_{i}\cdot\boldsymbol{q}}$,  $ e^{i\frac{1}{\sqrt{M}}P}$, and $ e^{\frac{1}{2} i \boldsymbol{R}_{i}\cdot\boldsymbol{p} }$. 
We will use the quadratic boundary actions,  
\begin{flalign}
S[Q] & =\frac{1}{K\pi}T\sum_{\omega_{n}} e^{-|\omega_n|/D} |\omega_{n}| |Q(\omega_{n})|^{2} +E_{C}\int d\tau(\frac{1}{\pi}\sqrt{M}Q-N_{g})^{2} \,,\quad S[\boldsymbol{q}]  =\frac{1}{K\pi}T\sum_{\omega_{n}} e^{-|\omega_n|/D} |\omega_{n}| |\boldsymbol{q}(\omega_{n})|^{2} \,, \\
 S[P] & = \frac{K}{4\pi}T \sum_{\omega_{n}} e^{-|\omega_n|/D}   \frac{|\omega_{n}|^2}{|\omega_{n}|+\frac{KM}{\pi}E_{C}}|P(\omega_{n})|^{2} \,, \quad  S[\boldsymbol{p}] = \frac{K}{4\pi} T \sum_{\omega_{n}} e^{-|\omega_n|/D} |\omega_{n}| |\boldsymbol{p}(\omega_{n})|^{2} \,. \label{eq:SPVertex}
\end{flalign}
The average of a vertex operator is non-negligible only for $ e^{2i\frac{1}{\sqrt{M}}Q}$ in which case the charging energy $E_C$ cuts off long-wave length fluctuations~\cite{PhysRevB.51.1743,matveev1994charge}. 
Since our Hamiltonian is time-independent, the two-point functions only depend on time-differences while the one-point function is time-independent. 

\subsection{The average \textmd{\normalsize{}$\left\langle e^{2i\frac{1}{\sqrt{M}} Q}\right\rangle$} 
\label{subsec:CorrFnQ}} 
For the average $\left\langle e^{2i\frac{1}{\sqrt{M}}Q}\right\rangle$ we find to lowest order in $T/E_C$ and $E_C/D$, 
\begin{flalign}
    \left\langle e^{2i\frac{1}{\sqrt{M}}Q(\tau)}\right\rangle & =e^{i2\frac{1}{M}\pi N_{g}}  \exp -\frac{K\pi T}{M}\sum_{\omega_{n}\neq0}e^{-|\omega_n|/D} \frac{1}{|\omega_{n}|+\frac{1}{\pi}KME_{C}} \\
    & =e^{i2\frac{1}{M}\pi N_{g}}  \left(\frac{Ke^{\gamma}ME_{C}}{\pi D}\right)^{K/M} \,. 
\end{flalign}
This result for $K=2$ was used in Sec.~\ref{subsec:1Strong-tunneling-limit} and for $K=1$ in Sec.~\ref{subsec:2Strong-tunneling-limit}.

\subsection{The correlation functions \textmd{\normalsize{}$\left\langle  e^{i\boldsymbol{R}_{i}\cdot[\boldsymbol{q}(\tau) - \boldsymbol{q}(0)]}\right\rangle $} and \textmd{\normalsize{}$\left\langle  e^{i\frac{1}{2}\boldsymbol{R}_{i}\cdot[\boldsymbol{p}(\tau) - \boldsymbol{p}(0)]}\right\rangle $} \label{subsec:CorrFnqandp}}

Correlation functions of type  $\langle  e^{i\sum_i  \boldsymbol{R}_{i}\cdot\boldsymbol{q}(\tau_i)}\rangle $ and $\langle  e^{i\frac{1}{2}\sum_i  \boldsymbol{R}_{i}\cdot\boldsymbol{p}(\tau_i)}\rangle $
are non-vanishing only when $ \sum_i  \boldsymbol{R}_{i} =0$. 
For the two-point function we find 
\begin{flalign}
\left\langle  e^{i\boldsymbol{R}_{i}\cdot[\boldsymbol{q}(\tau) - \boldsymbol{q}(0)]}\right\rangle & = \left\langle  e^{i T\sum_{\omega_{n}} \boldsymbol{B}(-\omega_{n})\cdot \boldsymbol{q}(\omega_{n}) } \right\rangle \\
& =\exp-\frac{K\pi}{4}T\sum_{\omega_{n}}e^{-|\omega_{n}|/D}\frac{1}{|\omega_{n}|}\boldsymbol{B}(-\omega_{n})\cdot\boldsymbol{B}(\omega_{n}) \\
& =\exp -\frac{K}{2}|\boldsymbol{R}_{i}|^{2}T \pi \sum_{\omega_{n}}e^{-|\omega_{n}|/D}\frac{1}{|\omega_{n}|}(1-\cos\omega_{n}\tau) \\ 
& =  \left(1+\frac{D^2}{\pi^2 T^2}\sin^2 \pi T\tau\right)^{-K \frac{1}{4}|\boldsymbol{R}_{i}|^{2} } \,. \label{eq:Corrq}
\end{flalign}
where $\boldsymbol{B}(-\omega_{n}) =  \boldsymbol{R}_{i}\, [e^{i\omega_{n}\tau}-1]$ and we used  the following sum~\cite{Gradshteyn} ($\omega_n = 2\pi n T$) valid in the limit $D\gg T$, 
\begin{flalign}
&  T\pi\sum_{\omega_{n}}e^{-|\omega_{n}|/D} \frac{1-\cos\omega_{n}\tau}{|\omega_{n}|} = \frac{1}{2} \ln \left(1+\frac{D^2}{\pi^2 T^2}\sin^2 \pi T\tau\right) \label{eq:int0} \,.
\end{flalign} 

The correlation function $\left\langle  e^{i\frac{1}{2}\boldsymbol{R}_{i}\cdot[\boldsymbol{p}(\tau) - \boldsymbol{p}(0)]}\right\rangle $ is calculated similarly. We find, 
\begin{equation}
    \left\langle  e^{i\frac{1}{2}\boldsymbol{R}_{i}\cdot[\boldsymbol{p}(\tau) - \boldsymbol{p}(0)]}\right\rangle  
    =  \left(1+\frac{D^2}{\pi^2 T^2}\sin^2 \pi T\tau\right)^{ -\frac{1}{4K}|\boldsymbol{R}_{i}|^{2}} \,. \label{eq:Corrp}
\end{equation}

From the correlation function~(\ref{eq:Corrq})  we find in the limit $T^{-1} \gg \tau \gg D^{-1}$ the result $\left\langle  e^{i\boldsymbol{R}_{i}\cdot[\boldsymbol{q}(\tau) - \boldsymbol{q}(0)]}\right\rangle \sim |\tau |^{- 2 \Delta}$, from which we can identify the scaling dimension $\Delta = K  \frac{1}{4}|\boldsymbol{R}_{i}|^{2}$.  
Similarly, from Eq.~(\ref{eq:Corrp}) we find $  \Delta = K^{-1}\frac{1}{4}|\boldsymbol{R}_{i}|^2$ for the operator $e^{i\frac{1}{2}\boldsymbol{R}_{i}\cdot \boldsymbol{p}} $. 
For the operator $e^{i\frac{1}{2}(\boldsymbol{R}_{i} -\boldsymbol{R}_{j} )\cdot \boldsymbol{p}} $, used in Eqs.~(\ref{eq:SStrongSCvalley}) and (\ref{eq:Tunneling_mu}), we have instead $\Delta = 2K^{-1}$.

\subsection{The correlation function \textmd{\normalsize{}$\left\langle e^{i\frac{1}{\sqrt{M}}[P(\tau)-P(0)]}\right\rangle $\label{subsec:The-correlation-function}}}

Let us  calculate the correlator $\langle e^{i\frac{1}{\sqrt{M}}[P(\tau)-P(0)]} \rangle $ by using the action $S[P]$ given in Eq.~(\ref{eq:SPVertex}) above. 
We introduce $B(-\omega_{n})=\frac{1}{\sqrt{M}}[e^{i\omega_{n}\tau}-1]$
which allows us to write 
\begin{flalign}
\left\langle e^{i\frac{1}{\sqrt{M}}[P(\tau)-P(0)]}\right\rangle_P & =\left\langle e^{Ti\sum_{i\omega_{n}}B(-\omega_{n})P(\omega_{n})}\right\rangle_P \\
& =\exp -\frac{\pi T}{2}\sum_{i\omega_{n}}e^{-|\omega_{n}|/D} \frac{|\omega_{n}|+\frac{2M}{\pi}E_{C}}{|\omega_{n}|^{2}}\left|\frac{1}{\sqrt{M}}(e^{i\omega_{n}\tau}-1)\right|^{2} \\
 & \approx\left(\frac{1}{\sqrt{1+  D^{2}\tau^{2}}}\right)^{1/M}e^{-E_{C}|\tau|} \,, \quad (T\ll\tau^{-1},\,E_{C},\,D)\,. \label{eq:CorrW=0}
\end{flalign}
We used  the following sums~\cite{Gradshteyn} ($\omega_n = 2\pi n T$) valid in the limit $D\gg T$, 
\begin{flalign}
&  T\pi\sum_{i\omega_{n}}e^{-|\omega_{n}|/D} \frac{1-\cos\omega_{n}\tau}{|\omega_{n}|} = \frac{1}{2} \ln \left(1+\frac{D^2}{\pi^2 T^2}\sin^2 \pi T\tau\right),\quad T\pi\sum_{i\omega_{n}}e^{-|\omega_{n}|/D}\frac{1-\cos\omega_{n}\tau}{|\omega_{n}|^2} = \frac{\pi}{2}  |\tau|(1+T|\tau|)  \label{eq:int} \,.
\end{flalign}

Equation~(\ref{eq:CorrW=0}) gives the correlation function when $N_{g}$ is close to an integer, e.g.,  $N_{g} \approx 0$. For half-integer values, $N_{g}\approx1/2$, we need to include the sum over winding number $W$ and consider instead 
\begin{equation}
C(\tau) = \left\langle e^{i\frac{1}{\sqrt{M}}[P(\tau)-P(0)]+2\pi iWT|\tau|}\right\rangle_{P,W} = \left\langle e^{i\frac{1}{\sqrt{M}}[P(\tau)-P(0)]}\right\rangle_{P}  \left\langle e^{2\pi i WT|\tau|}\right\rangle_{W}  \,,
\end{equation}
where the average over $W$ is evaluated with the help of $Z_0(N_g)$ in Eq.~(\ref{eq:Zp}).  
We find 
\begin{flalign}
\left\langle e^{2\pi iWT\tau}\right\rangle _{W} & =Z_{0}(N_{g})^{-1}\sum_{W=-\infty}^{\infty}e^{iN_{g}2\pi W-\pi^{2}TE_{C}^{-1}W^{2}}e^{i2\pi WT\tau}\\
 & =Z_{0}(N_{g})^{-1}\vartheta(N_{g}+T\tau;\,i\pi TE_{C}^{-1}) \,,
\end{flalign}
where $Z_{0}(N_{g})=\vartheta(N_{g};\,i\pi TE_{C}^{-1})$. At large
$E_{C}/T$, we can use the asymptotic  
\begin{equation}
Z_{0}(N_{g})^{-1}\vartheta(N_{g}+T\tau;\,i\pi TE_{C}^{-1})
\approx e^{-2N_{g}E_{C}|\tau|}\,.
\end{equation}
We find then  Eq.~(\ref{eq:Ctau}) of the main text: 
\begin{equation}
C(\tau) \equiv \left\langle e^{i\frac{1}{\sqrt{M}}[P(\tau)-P(0)]+2\pi iWT|\tau|}\right\rangle_{P,W} \approx\left(\frac{1}{\sqrt{1+D^2 \tau^{2}}}\right)^{1/M}e^{-E_{C}^{*}|\tau|}
= \begin{cases}
\left(\frac{1}{D \tau} \right)^{1/M}\,, \quad  1/D \ll \tau\,, 1/E_{C}^{*} \\ e^{-E_{C}^{*}|\tau|} \,, \quad \tau \ll 1/D \ll 1/E_{C}^{*}
\end{cases}
\,,
\end{equation}
where $E_{C}^{*}=2E_{C}(\frac{1}{2}-N_{g})$. 


\end{widetext}

 \section{Average charge \label{sec:Average-charge} \label{subsec:Ground-state-energy}}

We have from Eqs.~(\ref{eq:Hfull}) the charge
operator 
\begin{equation}
N=-\frac{1}{2E_{C}}\frac{dH}{dN_{g}}+N_{g}\,. \label{eq:ChargeDerivative}
\end{equation}
For calculating $\left\langle N\right\rangle $, it is therefore convenient
to find the $N_{g}$-dependent part of the average energy $\left\langle H\right\rangle $. 

\subsection{Weak tunneling limit \label{sec:Average-chargeWT}}

Let us first derive the average charge in the absence of tunneling. 
It is of course very easily obtained from the charging energy term of the Hamiltonian~(\ref{eq:Hfull}) and one finds $ \langle N \rangle =    \floor*{N_{g}+\frac{1}{2}} $  where $\floor*{\dots } $ is the floor function. (The charge steps are 1$e$-periodic since the topological superconductor can host an odd number of electrons.)
This result is obtained from the partition function~(\ref{eq:Zp}) as we show below.  

We are interested in the $N_{g}$-dependent
part of the average energy, which is obtained as $\left\langle H\right\rangle =-Z^{-1}\partial_{\beta}Z=-\partial_{\beta}\ln Z$
in general. 
Ignoring the $N_{g}$-independent contributions, we find the average energy, 
\begin{equation}
\left\langle H\right\rangle =-\frac{\partial_{\beta}Z_{0}(N_{g})}{Z_{0}(N_{g})}  \,,
\end{equation}
where $\beta = 1/T$ and 
\begin{equation}
Z_{0}(N_{g})=\sum_{W=-\infty}^{\infty}e^{iN_{g}2\pi W-\pi^{2}TE_{C}^{-1}W^{2}}=\vartheta(N_{g};\,i\pi TE_{C}^{-1}) \,.
\end{equation}
 For $E_{C}\gg T$, we can use the expansion of the theta function 
\begin{equation}
\vartheta(N_{g};\,i\pi T E_{C}^{-1})\approx\sqrt{\frac{E_{C}}{T \pi}}\exp-(E_{C}/T)(\{N_{g}-\frac{1}{2}\}-\frac{1}{2})^{2} \,,
\end{equation}
where $\{\dots\}$ is the positive fractional part. Using it, we find
\begin{flalign}
\langle H \rangle  &=-\partial_{\beta}\ln\vartheta(N_{g};\,i\pi TE_{C}^{-1}) \\
& \approx E_{C}(\{N_{g}-\frac{1}{2}\}-\frac{1}{2})^{2}\,,
\end{flalign}
for the ground state energy.  
Upon taking the derivative, we obtain 
$ \langle N \rangle = N_g -(\{N_{g}-\frac{1}{2}\}-\frac{1}{2})=   \floor*{N_{g}+\frac{1}{2}} $. 
Next, we will calculate the leading correction to this result due to weak tunneling between the island and the leads. 

The leading correction to ground state energy comes from a process where an electron tunnels through the junction $i$ into the island (or the lead)  and returns through the same junction within a time $\sim 1/ E_C$. 
The correction to charge due to finite tunneling is most convenient to calculate in the fermionic formalism, by using the tunneling action~(\ref{eq:StFermion}). 
This action corresponding to tunneling into a Majorana mode is valid in the limit of strong superconductor, $\Delta_P \gg E_C$. We will comment on  the opposite case below.  

From the tunneling action~(\ref{eq:StFermion}) in the Hamiltonian formalism, we find the 2nd order correction to ground state energy, 
\begin{equation}
    \delta E_{gs}^{(2)} \approx- \nu  \sum_{i}(Dt_{i})^{2}(\ln\frac{D}{\Delta E_{+1}}+\ln\frac{D}{\Delta E_{-1}}) \,, \label{eq:GSEnergy2}
\end{equation}
where $\nu \sim 1/D$ is the density of states (inverse bandwidth) of the lead and we denote $\Delta E_{\pm 1} $ the energy required to add/remove an electron from the island. For example, in the $N_g \approx 0$ valley we have $\Delta E_{\pm 1} =E_{C} (1  \mp 2N_{g})$. 
Upon taking the $N_g$-derivative of Eq.~(\ref{eq:GSEnergy2}), we find the leading  correction to the average charge 
\begin{equation} 
 \delta  \langle N \rangle^{(2)}	=\sum_{i} \frac{g_{i} \Delta_{P}}{E_C}  \frac{N_{g}}{(\frac{1}{4}-N_{g}^{2})} \,, \label{eq:N2}
\end{equation}
in the valley $-1/2 < N_g < 1/2$. 
We also  set the bandwidth $D\sim E_C$ and used the running coupling $t_{i}(D) \sim \sqrt{g_{i}}\sqrt{\Delta_{P}/D}$ from Eq.~(\ref{eq:tiRunning}). 
Equations~(\ref{eq:GSEnergy2})--(\ref{eq:N2}) become invalid in the narrow regions where $N_g$ is within $\sim M g \Delta_P / E_C $ of a charge degeneracy point, in which case for example $ \delta \langle N \rangle^{(2)}$ becomes of order 1. 
This estimate of the  regime of strong charge fluctuations around
the charge degeneracy point agrees with the estimate~(\ref{eq:peaksize}) obtained in the main text. 
Equation~(\ref{eq:N2}) gives the  result used in  Sec.~\ref{sec:StrongSC_charge} in the strong superconductor limit, $\Delta_P \gg E_C$.   

In the case of weak superconductivity ($\Delta_P \ll E_C$), the above estimates $\delta E_{gs}^{(2)}$ and $\langle N \rangle^{(2)}$  change as one needs to sum over intermediate energies of the virtual quasiparticles in the island, see Eq.~(\ref{eq:StFermionN}). 
This leads then to the result $ \delta  \langle N \rangle^{(2)} \sim \sum_{i} g_{i}  \ln\frac{1+2N_{g}}{1-2N_{g}} $ obtained in Ref.~\onlinecite{matveev1991quantum}.

\subsection{Strong tunneling limit \label{sec:Average-chargeST}}

In this Appendix we derive the correction to ground state charge from weak backscattering at the junctions. Our discussion is similar to Ref.~\onlinecite{aleiner2002quantum} where  a normal state dot is considered ($\Delta_P \to 0 $). 
In the absence of any backscattering ($r_i =0$) we have $\langle N \rangle =N_{g}$.

In order to calculate the leading-order backscattering correction to $\langle N \rangle$, we can calculate the correction to ground state energy by using $\left\langle H\right\rangle =-Z^{-1}\partial_{\beta}Z$ and then take the $N_g$-derivative as in Eq.~(\ref{eq:ChargeDerivative}). 
Upon integrating out the total charge mode $Q$, we obtain an effective action 
with a backscattering perturbation 
\begin{equation} \label{eq:SrApp}
  S_r =   E_C  \negthinspace\int\negthinspace d\tau\sum_{j}r_{j}(E_C)\cos(\mathbf{R}_{j}\cdot\boldsymbol{q}+2\frac{1}{M}\pi N_{g})\,,
\end{equation}
where we took the bandwidth $D \sim E_C$. 
This result is valid in both cases of strong ($\Delta_P \gg E_C$) and weak ($\Delta_P \ll E_C$) superconductivity. 
In the former limit it is obtained from  the action~(\ref{eq:Sfull}) and in the latter limit from Eq.~(\ref{eq:SfullWeakSC}). We have respectively $r_j(E_C) \sim (E_C/\Delta_P)\sqrt{1-g_j}$ and $r_j(E_C) \sim\sqrt{1-g_j}$ in Eq.~(\ref{eq:SrApp}), in the limit $g_j \approx 1$.

Upon expanding the partition function perturbatively in $S_r$, the leading term is in 2nd order $\propto r_j^2$. However, this contribution is $N_g$-independent and thus does not modify $\langle N \rangle $. 
The leading $N_g$-dependent term is in order $M$ perturbation theory $Z^{(M)} \sim  \left\langle \prod_{j=1}^{M} E_C \int d\tau_{j}r_{j}\cos[\mathbf{R}_{j}\cdot\boldsymbol{q}(\tau_{j})+2\frac{1}{M}\pi N_{g}]\right\rangle $ 
which is non-vanishing since $\sum_{j}\mathbf{R}_{j}=0$. 
The integrals  over  imaginary-time can be easily calculated by using Eq.~(\ref{eq:Corrq}). The integrals only depend on the  differences and are UV-divergent when $\tau_{i} - \tau_{j} \to 0$. This integration then yields $Z^{(M)} \sim (E_C/T) (\prod_j r_j(E_C)) \cos 2 \pi N_g$ and the correction to ground state energy 
$\delta E_{gs}^{(M)}\propto E_C  (\prod_j r_j (E_C)) \cos2\pi N_{g}$. 
By taking the $N_g$ derivative and substituting $r_j(E_C) \sim (E_C/\Delta_P)\sqrt{1-g_j}$ in the limit $\Delta_P \gg E_C$, 
we obtain the correction to $\langle N \rangle $ introduced in the main text, Eq.~(\ref{eq:StrongSCStrongTuncharge2}) [Eq.~(\ref{eq:StrongSCStrongTuncharge1}) is obtained similarly]. 
In the limit of weak superconductivity we substitute 
$r_j(E_C) \sim\sqrt{1-g_j}$ and find 
Eq.~(\ref{eq:WeakSCStrongTuncharge}).

\section{Current operator and Kubo formula for conductance \label{sec:currentop}}
In this Appendix we outline how the linear conductance presented in Sec.~\ref{subsec:StrongSCAverage-charge-and} is calculated by using the Kubo formula. We follow closely Ref.~\onlinecite{1995PhRvB..5216676F}.

The dc conductance matrix is given by~\cite{Mahan}
\begin{equation}
G_{ij}=i\lim_{\omega\to0}\omega^{-1}\lim_{i\omega_{n}\to\omega+i0^{+}}\int d\tau e^{i\omega_{n}\tau}\left\langle T_{\tau}I_{i}(\tau)I_{j}(0)\right\rangle \,, \label{eq:Kubo}
\end{equation}
where $I_i =-e\frac{1}{\pi}\partial_{t}\varphi_{i}$ is the current operator for junction $i$. 
In the two sections below, we use the Kubo formula to evaluate the conductance at low temperatures $T \ll E_C$. 
 
\subsection{Weak tunneling limit \label{sec:currentopWeak}}

In the Coulomb valley, the tunneling action is given by Eq.~(\ref{eq:SStrongSCvalley}), corresponding to electron tunneling from lead $i$ to lead $j$ without changing the charge of the island. 
By using the equation of motion for $\varphi_i$, we obtain the current operator 
\begin{equation}
I_{i} =-e\frac{1}{2}D\sum_{j}\lambda_{ij}(D)\sin\frac{1}{2}[\boldsymbol{R}_{i}-\boldsymbol{R}_{j}]\cdot\boldsymbol{p} \,.
\end{equation}
By using the current operator in the Kubo formula~(\ref{eq:Kubo}), we find 
\begin{widetext}

\begin{equation}
G_{ij}=i(e\frac{1}{2}D)^{2}\sum_{i'\neq i,j'\neq j}\lambda_{ii'}(D)\lambda_{jj'}(D)\lim_{\omega\to0}\omega^{-1}\lim_{i\omega_{n}\to\omega+i0^{+}}\int d\tau e^{i\omega_{n}\tau}\left\langle T_{\tau}\sin\frac{1}{2}[\boldsymbol{R}_{i}-\boldsymbol{R}_{i'}]\cdot\boldsymbol{p}(\tau)\sin\frac{1}{2}[\boldsymbol{R}_{j}-\boldsymbol{R}_{j'}]\cdot\boldsymbol{p}(0)\right\rangle 
\end{equation}
We have 
\begin{flalign}
\left\langle T_{\tau}\sin\frac{1}{2}[\boldsymbol{R}_{i}-\boldsymbol{R}_{i'}]\cdot\boldsymbol{p}(\tau)\sin\frac{1}{2}[\boldsymbol{R}_{j}-\boldsymbol{R}_{j'}]\cdot\boldsymbol{p}(0)\right\rangle  & =\frac{1}{(2i)^{2}}\sum_{\sigma\sigma'}\left\langle T_{\tau}e^{i\sigma\frac{1}{2}[\boldsymbol{R}_{i}-\boldsymbol{R}_{i'}]\cdot\boldsymbol{p}(\tau)}e^{i\sigma'\frac{1}{2}[\boldsymbol{R}_{j}-\boldsymbol{R}_{j'}]\cdot\boldsymbol{p}(0)}\right\rangle \\
& =-\frac{1}{2}\left(\delta_{i'j}\delta_{ij'}+\delta_{ij}\delta_{i'j'}\right)\left\langle T_{\tau}e^{i\frac{1}{2}[\boldsymbol{R}_{j}-\boldsymbol{R}_{j'}]\cdot\boldsymbol{p}(\tau)}e^{-i\frac{1}{2}[\boldsymbol{R}_{j}-\boldsymbol{R}_{j'}]\cdot\boldsymbol{p}(0)}\right\rangle 
\end{flalign}
and 
\begin{equation}
\left\langle T_{\tau}e^{i\frac{1}{2}[\boldsymbol{R}_{j}-\boldsymbol{R}_{j'}]\cdot\boldsymbol{p}(\tau)}e^{-i\frac{1}{2}[\boldsymbol{R}_{j}-\boldsymbol{R}_{j'}]\cdot\boldsymbol{p}(0)}\right\rangle =\left(\frac{D}{\pi T}\sin\pi T\tau\right)^{-K^{-1} \frac{1}{2}|\boldsymbol{R}_{j}-\boldsymbol{R}_{j'}|^{2}} \,.
\end{equation}
where $K=2$ in the superconducting case and $K=1$ in the normal
case, and $\ensuremath{\frac{1}{4}|\boldsymbol{R}_{j'}-\boldsymbol{R}_{j}|^{2}=2}$
for $j\neq j'$. Thus, $\Delta=1$ and $\Delta=2$ in the superconducting
and normal cases.
Thus 
\begin{equation}
G_{ij}=-i(-e\frac{1}{2}D)^{2}\lim_{\omega\to0}\omega^{-1}\lim_{i\omega_{n}\to\omega+i0^{+}}\frac{1}{2}\left(\lambda_{ij}(D)^{2}[1-\delta_{ij}]-\delta_{ij}\sum_{i'\neq i}\lambda_{ii'}(D)^{2}\right)\int d\tau e^{i\omega_{n}\tau}\left(\frac{D}{\pi T}\sin\pi T\tau\right)^{-4K^{-1}} \,.
\end{equation}
We take the real part and use the integral in the limit $\omega_n \to 0$, 
\begin{equation}
\int_{D^{-1}}^{T^{-1}-D^{-1}}d\tau\frac{1-\cos\omega_{n}\tau}{\left(\frac{D}{\pi T}\sin\pi T\tau\right)^{\nu}}  =-i\omega_{n}i\left(\frac{\pi T}{D}\right)^{\nu}\frac{1}{2\pi T^{2}}B(\frac{1}{2},\frac{\nu}{2})\,, \label{eq:Integral}
\end{equation}
where $B$ is the beta function. 
We find 
\begin{equation}
G_{ij}  =\frac{e^{2}}{h}\frac{\pi^{2}}{8}\left(\lambda_{ij}(D)^{2}[1-\delta_{ij}]-\delta_{ij}\sum_{i'\neq i}\lambda_{ii'}(D)^{2}\right)\left(\frac{\pi T}{D}\right)^{-2+4K^{-1}}B(\frac{1}{2},2K^{-1}) \,. \label{eq:AppGTunGeneral}
\end{equation}
\end{widetext}
In the superconducting case we have $\ensuremath{\lambda_{ij}(D)}\approx\sqrt{g_{i}g_{j}}\Delta_{P}/E_{C}^{*}$
with logarithmic corrections, see Sec.~\ref{subsec:StrongSCCoulomb-valley}. 
Thus, in the superconducting case ($K=2$), 
\begin{flalign}
G_{ij} & =\frac{e^{2}}{h}\frac{\pi^{2}}{4}g_{i}\left(g_{j}-\delta_{ij}g_{\Sigma}\right)\left(\frac{\Delta_{P}}{E_{C}^{*}}\right)^{2} \left[\frac{\ln (eE_C/T_K)  }{ \ln (eT / T_K)} \right]^2 \,, 
\end{flalign}
where we have included the logarithmic corrections. 
This is  Eq.~(\ref{eq:GStrongSCKondo}) of the main text. 
In the normal case ($K=1$), we have for $D\ll E_C^*$ the running coupling  $\ensuremath{\lambda_{ij}(D)\to \lambda_{ij}(D)\sqrt{g_{i}g_{j}}D/E_{C}^{*}}$, in the notation of  Sec.~\ref{subsec:WeakSCCoulomb-valley}. Thus, we find 
\begin{flalign}
G_{ij} & =\frac{e^{2}}{h}\frac{\pi^{2}}{6}g_{i}\left(g_{j}-\delta_{ij}g_{\Sigma}\right)\left(\frac{\pi T}{E_{C}^{*}}\right)^{2}\,, \label{eq:AppGTunNormal}
\end{flalign}
where $g_{\Sigma}=\sum_{i}g_{i}$. 

We can also use Eq.~(\ref{eq:AppGTunGeneral}) in the case of strong tunneling when $\Gamma \gg \Delta_P$ in which case tunneling becomes weak in the temperature interval $\Delta_P \ll T \ll \Gamma$. 
In this limit, we have $\lambda_{ij}(D)=\lambda_{ij}(\Gamma)(D/\Gamma)$, see discussion in Sec.~\ref{subsec:2Strong-tunneling-limit} above Eq.~(\ref{eq:TKWeakScTun}). Therefore, from  Eq.~(\ref{eq:AppGTunGeneral}) with $K=1$ we find,  
\begin{equation}
G_{ij}  = \frac{e^{2}}{h} \left( \lambda_{ij}(\Gamma)^2-\delta_{ij} \sum_k \lambda_{ik}(\Gamma)^2\right)\left(\frac{ T}{\Gamma}\right)^{2}, \label{eq:AppGTunGamma}
\end{equation}
which was given in Eq.~(\ref{eq:GTunGamma}) of the main text.

\subsection{Strong tunneling limit \label{sec:currentopStrong}}

In the strong tunneling limit, it is easiest to work in terms of the $\varphi$ variables. The current operator   is then  $I_{i}=-e\frac{1}{\pi}\partial_{t}\varphi_{i}$ and the Kubo formula becomes~(\ref{eq:Kubo}) 
\begin{equation}
    G_{ij}=i\frac{e^2 T}{\pi^2}\lim_{\omega\to0}\omega \lim_{i\omega_{n}\to\omega+i0^{+}}\left\langle \varphi_{i}(-i\omega_{n})\varphi_{j}(i\omega_{n})\right\rangle \,. \label{eq:Kubo2}
\end{equation}
This form for the conductance is best suited for calculations in the strong tunneling limit. 


We can write 
\begin{equation}
\left\langle \varphi_{i}(-i\omega_{n})\varphi_{j}(i\omega_{n})\right\rangle =\left.\frac{1}{Z_{J}}\frac{\delta}{\delta J_{i}(i\omega_{n})}\frac{\delta}{\delta J_{j}(-i\omega_{n})}Z_J \right|_{J=0}\,,
\end{equation}
where 
\begin{equation}
Z_{J}=\int\mathcal{D}\varphi_{j}e^{-S-S_{J}}\,,\,\, S_{J}=-\sum_{\omega_{n}}\sum_{j}J_{j}(-i\omega_{n})\varphi_{j}(i\omega_{n}).
\end{equation}

In the presence of charging energy, the proper variables are $\boldsymbol{q}$ and $Q$, Eq.~(\ref{eq:phiqbasis}). 
In terms of these variables we have 
\begin{widetext}

\begin{flalign}
S & =S_{0}-D\int d\tau\sum_{j}r_{j}(D) \cos(\mathbf{R}_{j}\cdot\boldsymbol{q}+2\frac{1}{\sqrt{M}}Q) \,, \label{eq:SAppendixBS1} \\
S_{0} & = \frac{1}{\pi}T\sum_{i\omega_{n}} \frac{1}{K}|\omega_{n}|\left(|\boldsymbol{q}(\omega_{n})|^{2}+|Q(\omega_{n})|^{2}\right)+E_{C}\int d\tau(\frac{1}{\pi}\sqrt{M}Q-N_{g})^{2} \,, \\
S_{J} & =-\sum_{i\omega_{n}}\sum_{j}J_{j}(-i\omega_{n})[\frac{1}{2}\boldsymbol{R}_{j}\cdot\boldsymbol{q}(i\omega_{n})+\frac{1}{\sqrt{M}}Q(i\omega_{n})] \,. \label{eq:SAppendixBS3}
\end{flalign} 
In order to describe both normal and superconducting islands, we
have included in $S_{0}$ a factor $K$ which is $1$ in the normal
case ($D \gg \Delta_P$) and $2$ in the topological superconducting case ($D\ll \Delta_P$). 
Likewise, in the former case the reflection amplitude  is irrelevant under RG flow, $r_i(D) =(D/{\Delta_P} )^{K -1} \sqrt{1-g_i} $, where  $E_C \ll D \ll \Delta_P$. 
The above action thus allows us to cover both actions~(\ref{eq:Sfull}) (when $\Delta_P \gg E_C$) and (\ref{eq:SfullWeakSC})  (when $\Delta_P \ll E_C$).

 Expanding $e^{-S-S_{J}}$
in perturbation theory in reflection, we find to second order in $r_{j}$, 
\begin{flalign}
e^{-S-S_{J}} & =e^{-S_{0}-S_{J}}\left(1+\frac{1}{2}D^{2}\int d\tau_{1}\int d\tau_{2}\sum_{j_{1}j_{2}}r_{j_{1}}r_{j_{2}}\cos[\mathbf{R}_{j_{1}}\cdot\boldsymbol{q}(\tau_{1})+2\frac{1}{\sqrt{M}}Q(\tau_{1})]\cos[\mathbf{R}_{j_{2}}\cdot\boldsymbol{q}(\tau_{2})+2\frac{1}{\sqrt{M}}Q(\tau_{2})]\right)\\
 & =e^{-S_{0}-S_{J}}\left(1+\frac{1}{2}D^{2}\int d\tau_{1}\int d\tau_{2}\sum_{j_{1}j_{2}}r_{j_{1}}r_{j_{2}}\frac{1}{4}\sum_{\sigma_{1},\sigma_{2}=\pm}e^{\sum_{i\omega_{n}}\mathbf{A}(-i\omega_{n})\cdot\boldsymbol{q}(i\omega_{n})}e^{\sum_{i\omega_{n}}B(-i\omega_{n})Q(i\omega_{n})}\right)
\end{flalign}
where 
\begin{equation}
\mathbf{A}(-i\omega_{n})=iT\left(\sigma_{1}e^{i\omega_{n}\tau_{1}}\mathbf{R}_{j_{1}}+\sigma_{2}e^{i\omega_{n}\tau_{2}}\mathbf{R}_{j_{2}}\right)\,,\quad B(-i\omega_{n})=2i\frac{1}{\sqrt{M}}T\left(\sigma_{1}e^{i\omega_{n}\tau_{1}}+\sigma_{2}e^{i\omega_{n}\tau_{2}}\right)
\end{equation}
Thus, 
\begin{flalign}
Z_{J} & =Z_{0}\left\langle e^{-S_{J}}\left(1+\frac{1}{2}D^{2}\int d\tau_{1}\int d\tau_{2}\sum_{j_{1}j_{2}}r_{j_{1}}r_{j_{2}}\frac{1}{4}\sum_{\sigma_{1},\sigma_{2}=\pm}e^{\sum_{i\omega_{n}}\mathbf{A}(-i\omega_{n})\cdot\boldsymbol{q}(i\omega_{n})}e^{\sum_{i\omega_{n}}B(-i\omega_{n})Q(i\omega_{n})}\right)\right\rangle _{S_{0}} \label{eq:ZJ2ndOrder}
\end{flalign}
The averages over $\boldsymbol{q}$ and $Q$ can be done separately.
For this, we have the following correlation functions, 
\begin{flalign}
\left\langle e^{\sum_{i\omega_{n}}\mathbf{C}(-i\omega_{n})\cdot\boldsymbol{q}(i\omega_{n})}\right\rangle _{S_{0}} & =e^{\frac{1}{2}K T^{-1}\frac{\pi}{2}\sum_{i\omega_{n}}|\omega_{n}|^{-1}\mathbf{C}(i\omega_{n})\cdot\mathbf{C}(-i\omega_{n})}\,,\\
\left\langle e^{\sum_{i\omega_{n}}C(-i\omega_{n})Q(i\omega_{n})}\right\rangle _{S_{0}} & =e^{\pi T^{-1}\frac{1}{2}\sum_{i\omega_{n}\neq0}C(-\omega_{n})\left(2K^{-1}|\omega_{n}|+\frac{2}{\pi}ME_{C}\right)^{-1}C(\omega_{n})}e^{\frac{1}{4}T^{-1}\frac{\pi^{2}}{ME_{C}}C(0)^{2}}e^{T^{-1}C(0)\frac{\pi N_{g}}{\sqrt{M}}}\,.
\end{flalign}

We find for $\omega_{n},\,T\ll E_{C}$, %
\begin{flalign}
& \left\langle \varphi_{i}(-i\omega_{n})\varphi_{j}(i\omega_{n})\right\rangle \nonumber  \\
&= \frac{\pi (\delta_{ij}-\frac{1}{M}) }{2K^{-1}T|\omega_{n}|}
+2\frac{\pi^{2}D^{2}}{4K^{-2}|\omega_{n}|^{2}T}\left(\frac{2e^{\gamma}ME_{C}}{2K^{-1}\pi D}\right)^{2K/M}
\left(\frac{D}{E_C} \right)^{2K -2}  R_{ij}
\int_{D^{-1}}^{T^{-1}-D^{-1}}d\tau\frac{1-\cos\omega_{n}\tau}{\left(\frac{D}{\pi T}\sin\pi T\tau\right)^{2K(1-\frac{1}{M})}}\,, \label{eq:phiphi}
\end{flalign}
where the factor $\left(\frac{D}{E_C} \right)^{2K -2}$ is absent in the normal case $K=1$ ($D \gg \Delta_P$). We denote
\begin{equation}
R_{ij}=-\sum_{j_{1}} r_j(E_C)^2 \left(\delta_{jj_{1}}-\frac{1}{M}\right)\left(\delta_{ij_{1}}-\frac{1}{M}\right) \,, \label{eq:Rij}
\end{equation}
where $ r_j(E_C)^2 = \left(\frac{E_C}{\Delta_P} \right)^{2K -2} (1-g_{j_{1}})$. 
For roughly isotropic contacts, we have $R_{ij} \approx r(E_C)^2  \left(\frac{1}{M}- \delta_{ij}\right)$. 
We also used the following results for $\sigma_{2}=-\sigma_{1}$
and $j_{1}=j_{2}$ [which is the dominant contribution to $Z_J$,  Eq.~(\ref{eq:ZJ2ndOrder})], 
\begin{flalign}
\left\langle e^{\sum_{i\omega_{n}}B(-i\omega_{n})Q(i\omega_{n})}\right\rangle _{Q} & =\left\langle e^{2i\frac{1}{\sqrt{M}}\sigma_{1}\left(Q(\tau_{1})-Q(\tau_{2})\right)}\right\rangle _{Q}\approx\left(\frac{2e^{\gamma}ME_{C}}{g\pi D}\right)^{2K/M}\,.
\end{flalign}
and 
\begin{flalign}
e^{\frac{K}{2}T^{-1}\frac{\pi}{2}\sum_{i\omega_{n}}|\omega_{n}|^{-1}\mathbf{A}(i\omega_{n})\cdot\mathbf{A}(-i\omega_{n})} & =e^{-(1-\frac{1}{M})\frac{1}{2}K T\pi\sum_{i\omega_{n}}|\omega_{n}|^{-1}[1-\cos\omega_{n}(\tau_{1}-\tau_{2})]}\\
 & =\left(\frac{D}{\pi T}\sin\pi T(\tau_{1}-\tau_{2})\right)^{-4(1-\frac{1}{M})\frac{1}{2}K}\,.
\end{flalign}
The last integral in Eq.~(\ref{eq:phiphi}) in the limit $\omega_{n}\to0$ can be calculated by using the integral~(\ref{eq:Integral}) with $\nu = 2K(1-\frac{1}{M})$. %
Thus, with the help of Eq.~(\ref{eq:Kubo2}) we obtain 
\begin{equation}
G_{ij}/G_{0}=K\left(\frac{1}{M} - \delta_{ij}\right)-\frac{\pi^{2}K^{2}}{2}
\left(\frac{\pi T}{E_C} \right)^{ 2K -2 } 
\left(\frac{Ke^{\gamma}ME_{C}}{ \pi^2 T}\right)^{2K/M}
 B\! \left(\frac{1}{2},K \left(1-\frac{1}{M}\right)\right)     R_{ij}\,.
\end{equation}
where $G_{0}=e^{2}/h$.
This result covers both the normal ($T \gg \Delta_P$)  and superconducting ($T \ll \Delta_P$) cases by setting the parameter $K = 1$ or 2, respectively. 
In the superconducting case, $K=2$, we obtain Eq.~(\ref{eq:GStrongTun}) of the main text. 
In the normal case, $T\gg\Delta_{P}$, i.e., $K=1$, we find 
\begin{equation}
G_{ij}/G_{0}=\left(\frac{1}{M} - \delta_{ij}\right)-\frac{1}{2}\pi^{2}\left(\frac{e^{\gamma}ME_{C}}{\pi^{2}T}\right)^{2\frac{1}{M}}
 B\! \left(\frac{1}{2},\left(1-\frac{1}{M}\right)\right)    R_{ij}\,. \label{eq:GStrongTunNormal}
\end{equation}
This is Eq.~(\ref{eq:GWeakSCStrongTunAboveGap}) of the main text. 
In that equation, we have $b_M=\frac{1}{2}\pi^{2} \left(\frac{e^{\gamma}M}{\pi^{2}}\right)^{2/M} B\! \left(\frac{1}{2},\left(1-\frac{1}{M}\right)\right)$; $B$ is the beta function and 
$8.5 \leq b_M \leq 11.4$ for $M=3,\dots,6$. 
For $M=4$, the temperature-dependence agrees with the result of two spinful leads in Ref.~\onlinecite{1995PhRvB..5216676F}.  
Equation~(\ref{eq:GStrongTunNormal}) is valid at $E_C \gg T \gtrsim \Delta_P$ and the $T$-dependence arises from the renormalization of reflection amplitudes to scale $T$, see below Eq.~(\ref{eq:ScalingDimReflWSC}). 
At $T \ll \Delta_P$, we can replace $T\to \Delta_{P}$ in Eq.~(\ref{eq:GStrongTunNormal}) and multiply the second term by a factor $(T/\Delta_{P})^{2(1-\frac{2}{M})}$ as  the reflection amplitudes get renormalized further.

We can also use the above method to evaluate the backscattering correction to conductance near the strong-coupling fixed point, $T \ll T_K$. 
We replaces Eqs.~(\ref{eq:SAppendixBS1})--(\ref{eq:SAppendixBS3}) by 
\begin{equation}
S  =\frac{1}{2\pi}T\sum_{\omega_{n}}e^{-|\omega_n|/T_K} |\omega_{n}||\boldsymbol{q}(\omega_{n})|^{2} - T_K 
   \negthinspace\int\negthinspace d\tau\sum_{j}r_{j}(T_K)\cos(\mathbf{R}_{j}\cdot\boldsymbol{q}+2\frac{1}{M}\pi N_{g})\,, \label{eq:StrongCouplingBS}
\end{equation}
where $T_K$ is now the UV cutoff and $r_j (T_K) \sim 1$.  
Instead of Eq.~(\ref{eq:phiphi}) we now find (for $\omega_{n},\,T\ll T_{K}$), 
\begin{flalign}
& \left\langle \varphi_{i}(-i\omega_{n})\varphi_{j}(i\omega_{n})\right\rangle \nonumber  \\
&= \frac{\pi (\delta_{ij}-\frac{1}{M}) }{T|\omega_{n}|}
- 2\frac{\pi^{2}T_K^{2}}{|\omega_{n}|^{2}T} 
\sum_{j_{1}} r_{j_{1}}(T_K)^2 \left(\delta_{jj_{1}}-\frac{1}{M}\right)\left(\delta_{ij_{1}}-\frac{1}{M}\right)
\int_{T_K^{-1}}^{T^{-1}-T_K^{-1}}d\tau\frac{1-\cos\omega_{n}\tau}{\left(\frac{T_K}{\pi T}\sin\pi T\tau\right)^{4(1-\frac{1}{M})}}\,, \label{eq:phiphiBelowTK}
\end{flalign}
and the conductance 
\begin{equation}
    G_{ij} / G_0=   2 \left(\frac{1}{M} - \delta_{ij}\right)
+  2 \pi^2  B\! \left(\frac{1}{2},2\left(1-\frac{1}{M}\right)\right)   
\sum_{j_{1}} r_{j_{1}}(T_K)^2 \left(\delta_{jj_{1}}-\frac{1}{M}\right)\left(\delta_{ij_{1}}-\frac{1}{M}\right)
\left(\frac{\pi T}{T_K}\right)^{2(1-\frac{2}{M})}
 \,, \label{eq:GStrongCoupling}
\end{equation}
which gives Eq.~(\ref{eq:conductanceBelowTk}) of the main text upon setting $r_{j_{1}}(T_K)^2 \sim 1$ and $\sum_{j_{1}}  \left(\delta_{jj_{1}}-\frac{1}{M}\right)\left(\delta_{ij_{1}}-\frac{1}{M}\right) = -(\frac{1}{M} -\delta_{ij} )$. 
\end{widetext}

\bibliography{refs}

\end{document}